\documentclass[pra,twocolumn,twoside,aps]{revtex4-1}

\usepackage{amsmath, amssymb}
\usepackage{bm}
\usepackage{natbib}
\usepackage[dvipsnames]{xcolor}

\usepackage{graphicx}
\usepackage{float} 

\usepackage{graphicx}
\usepackage[colorlinks=true,linkcolor=blue,citecolor=red,urlcolor=blue]{hyperref}
\usepackage{amssymb}

\begin{document}

\title{A trillion frames per second: the techniques and applications of light-in-flight photography.}

\author{Daniele Faccio$^{1}$, Andreas Velten$^{2}$}
\address{$^{1}$School of Physics \& Astronomy, University of Glasgow, Glasgow G12 8QQ, UK.\\
$^{2}$Department of Biostatistics and Medical Informatics, University of Wisconsin, Madison, WI 53706, USA.}

\email{daniele.faccio@glasgow.ac.uk;velten@wisc.edu}


\begin{abstract}
Cameras capable of capturing videos at a trillion frames per second allow to freeze light in motion, a very counterintuitive capability when related to our everyday experience in which light appears to travel instantaneously. By combining this capability with computational imaging techniques, new imaging opportunities emerge such as three dimensional imaging of scenes that are hidden behind a corner, the study of relativistic distortion effects, imaging through diffusive media and imaging of ultrafast optical processes such as laser ablation, supercontinuum and plasma generation. We provide an overview of the main techniques that have been developed for ultra-high speed photography with a particular focus on `light-in-flight' imaging, i.e. applications where the key element is the imaging of light itself at frame rates that allow to freeze it's motion and therefore extract information that would otherwise be blurred out and lost. 
\end{abstract}


%
\maketitle

\tableofcontents

\section{Introduction.}
Images are everywhere and imaging, the art of taking ``good'' images is either the result or the driving force of huge investments in technology and research. Yet, although it was first demonstrated with analog film-photographic techniques, the recent increase in activity in this field was enabled by the invention of CCD imagers in the late `70s and their subsequent low-cost, mass production starting from the late `90's. Digital imaging with light-weight or cell-phone cameras has become a new vehicle of mass communication, supplanting more traditional forms of sharing information. All this has in turn driven new forms of processing the data contained in the images with innovations from the field of computational imaging reaching the market and consumer products in remarkably short times. Digital sensors have also allowed the fusion of what used to be two distinct imaging fields: still imaging and video recording. With one and the same technology/sensor we can decide to take either a still image or record a full video, including effects such as slow-motion recording. And this brings us to the focus of this review.\\
\indent The first photograph taken in Le Gras by Nic{\'e}phore Ni{\'e}pce in 1828 required several days of exposure and resulted in a rather grainy, indistinct black and white image. Yet, this feat laid the foundations for the current imaging revolution we are living today. The techniques introduced by Nic{\'e}phore and nearly simultaneously by Daguerre and Talbot where quickly developed and improved upon. In the late 1800's photography was a well developed technology and, although not yet generally accepted as an art form, it was highly considered for its factual qualities and ability to provide information and images from distant places and worlds that otherwise would remain unreachable. It is this objective nature of photography that made it the preferred vehicle for example for nature conservationists in promoting the cause to create the first national parks in the USA. But until the late 1880's photography was just this: a means to freeze and record, and therefore share with others, what the human eye could already see. This changed when Eadweard Muybridge, in the attempt to determine the exact dynamics of a galloping horse, invented a simple technique with which he could freeze the motion of events that were occurring so fast as to be invisible to the human eye. His idea was relatively simple: he placed a series of wire traps along the path of the horse, each one connected to a camera with a fast mechanical shutter that was triggered by the horse tripping the wire. In this way Eadweard recorded the now iconic sequence of images, ``The Horse in Motion'' (1878). For the first time, photography was used to capture something new, something that the eye on its own cannot see.\\
\indent As was the case with Nic{\'e}phore's first images, Eadweard's invention had plenty of space for improvement. Mechanical shutters and emulsion films were developed allowing exposures as short one thousandth of a second. However, such short exposures are usually unpractical due to the very low light levels incurred when using short exposure times.  Flash photography was the next development in the field: a bright flash of light provided at the same time a brought source of light, thus solving the problem of low light levels whilst ensuring a capture mechanism with times of the order of 1/1000 second.\\
\indent Harold Eugene Edgerton, working at MIT between 1934 and 1977, transformed what were then rudimental forms of active imaging into true art forms and exceptional scientific tools. Harold is generally acknowledged for the development of stroboscopic lighting that he used for scientific and artistic purposes alike, winning an Oscar in 1940 for his short ``Quicker'n a Wink''. He also popularised the air-gap flash, based on a high-voltage electrical discharge and originally invented by Talbot. The air-gap flash routinely achieves exposure times in the microsecond range, short enough to freeze the motion of a supersonic bullet and to generate some of Harold's most iconic images of bullet Mach cones. Harold also invented a completely new approach to high-speed photography by revisiting the concept of the shutter placed in front of the camera. His Raptronic camera used a magneto-optic cell placed in between two crossed polarisers. A short electrical current rotates the polarisation of light that has passed through the first polariser with exposure times on the camera as short as 10 nanoseconds. This camera was used for example to capture the first high speed images of a nuclear explosion with remarkable detail providing new insight into the explosion plasma dynamics. Similarly to the technique introduced by Muybridge, each camera could only take one still image at a time: in order to build a sequence of images, a series of synchronised cameras were used to image the same event.\\
\indent The methods described so far allow to capture up to several thousand frames a second, albeit limited to just several frames (due to the requirement to use one camera for each individual frame). However, modern commercially available individual cameras can provide full movies with frames up to staggering rates of the order of $8\cdot10^5$ fps or higher and exposure times down to 500 ns. This is all thanks to advances in CMOS technology that is gradually replacing CCD layouts in modern digital cameras. Whilst CCD cameras rely on a transfer of the photo-induced charges to a common buffer on the chip, in CMOS cameras each pixel can be read out individually. This allows, with the use of high sensitivity pixels, high speed read-out circuitry and large memory buffers, to obtain remarkable frame rates: the mechanical shutter is now being gradually replaced by the fast acquisition and read-out of the CMOS chip itself. Currently a trade-off needs to be met with such cameras that arises more from the read-out electronics and memory capacity rather than from the speed of the acquisition itself. Only a certain amount of data can be read-off the chip every second before either saturating the memory or the information capacity of the read-out electronics. Therefore, modern cameras will typically quote the camera speed in Giga-pixels per-second, rather than in frames per second. For example, a 1 Gpx/second camera can provide 12600 fps at 1280x800 pixel resolution or 820000 fps at 128x16 pixel resolution. \\
\indent In the following we will discuss a series of technologies aimed at collecting images with exposure times and frame rates that are orders of magnitude higher than those achievable with the cameras discussed so far. 
Exposure times of the order of 1 ps and frame rates of 1 trillion frames per second are possible opening up exciting possibilities such as capturing the motion not of supersonic bullets but of actual pulses or bullets of light.
Such frame rates can be achieved by different approaches, each with its own advantages and drawbacks and should therefore be chosen depending on the application. Recent successes of these technologies include for example capturing the motion of light-in-flight (LiF), three dimensional reconstruction of scenes hidden from view, motion tracking of objects from behind a corner and the observation of relativistic effects due to the finite propagation speed of light itself, i.e. the camera is so fast that it can detect ``aberrations'' arising from the travel time of light from the scene to the camera.\\
We distinguish LiF technology from the more commonly known Time-of-Flight (ToF) techniques. ToF refers to a multitude of applications that rely on measurements of the time taken for a pulse of light to travel, reflect off an object and return to the sender. The main example is LIDAR ranging and 3D scene reconstruction, that comes in a variety of implementations. The two most common approaches rely on either scanning the scene with a laser spot and collecting the return signal (usually confocally, i.e. along the same line-of-sight of the laser beam) or flash-illuminating the whole scene with a broad beam and then using a detector array or time-resolving camera to capture the return times from different points simultaneously. Much of the technology used for LIDAR and related applications is common to LiF photography, thus allowing in principle to convert from one approach to the other.  However,  while LIDAR and also standard cameras traditionally use time to create 3D renditions of the scene, with LiF photography we now capture the motion of light itself, which is more a part of the imaging process rather than of the scene itself. Light-in-Flight is sometimes also referred to as `transient imaging', in particular in the context of computer graphics and vision. Jarabo et al. have prepared a recent review of the computational methods adopted for LiF or transient imaging \cite{jarabo} and Bhandari et al. provide a tutorial overview for the more general problem of time-of-flight imaging \cite{raskar_tutorial}. In this review we will focus attention more on the available illumination/detection technologies and applications of LiF, with only a brief overview of some of the computational retrieval techniques. \\
\begin{figure}[t]
\includegraphics[width=8cm]{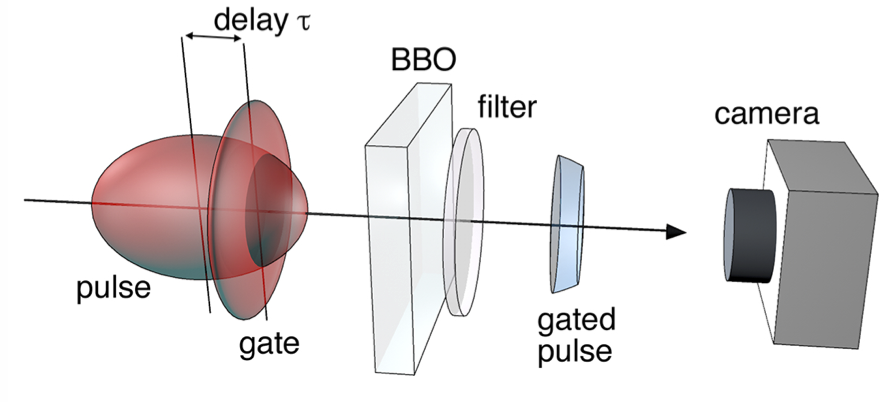}
\begin{center}
\caption{Light-in-flight by nonlinear optical gating: the figures shows one possible layout of a nonlinear optical gating scheme where a very short gate laser pulse (e.g. 10-20 fs temporal duration) is overlapped with the longer laser pulse that one wishes to measure. This laser pulse may be, for example, a pulse that has undergone complex transformation after propagation in a nonlinear medium, as shown in Fig.~\ref{bullet} below. The pulse and gate overlap inside a nonlinear crystal (BBO) that converts the frequency (e.g. through sum-frequency generation) only at the overlap. A filter placed after the crystal allows only the sum-frequency signal to pass and to be finally recorded on a camera. Measurements are repeated with a varying temporal delay, $\tau$ so as to reconstruct the pulse slice-by-slice.}
\label{pump-probe}
\end{center}
\end{figure}
\begin{figure*}[t]
\includegraphics[width=16cm]{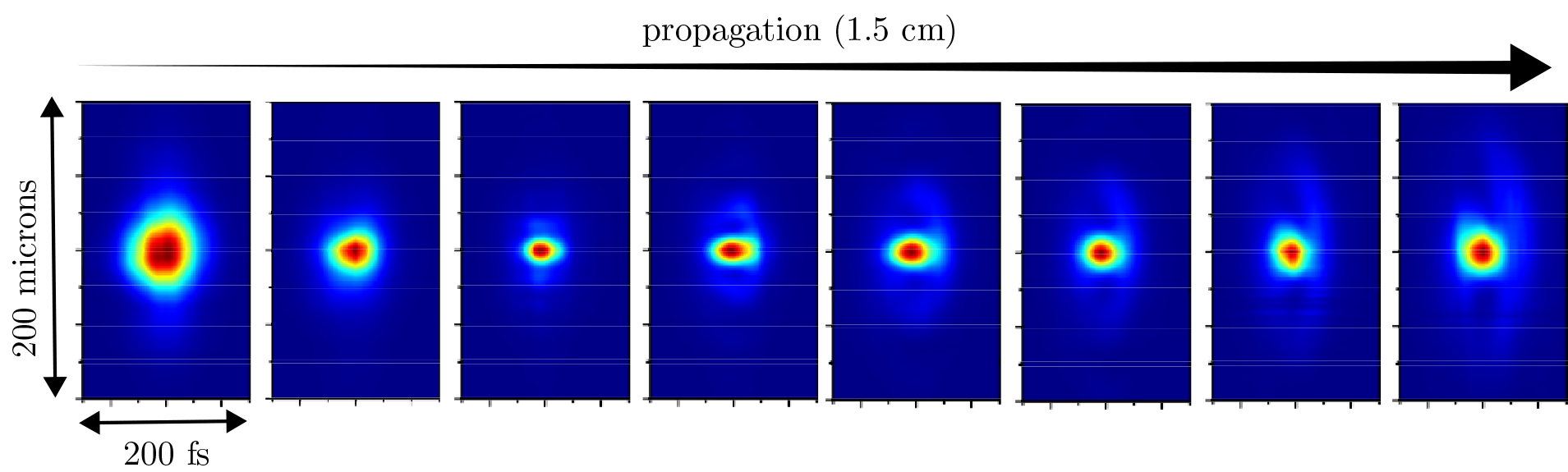}
\begin{center}
\caption{Light-in-flight video at 0.1 trillion frames/second by nonlinear optical gating: experimental results showing the propagation and evolution of a high intensity light bullet propagating through glass. Each frame corresponds to a different propagation position through the glass, equally distributed along the total 1.5 cm distance. The full LiF tomographic reconstruction of the light pulse shown here allows to infer details on the interaction mechanisms of the intense light pulse with the medium, that in turn lead to the observed reshaping during propagation. Figure adapted from Ref.\cite{SH5}.}
\label{bullet}
\end{center}
\end{figure*}
\section{Capturing light-in-flight.}
With the advent of reliable pulsed lasers in the 1970's, a natural desire emerged to follow the transient dynamics of a short pulse of light, both as a curiosity but also in relation to specific applications that relied on these transients and whose dynamics rely on the ability to control the laser pulse behaviour on extremely short time scales, from nanoseconds all the way down to femtoseconds. For example, one of the first applications of pulsed lasers was material processing and in particular pulsed laser deposition of various materials. Other applications involved the use of short pulses to probe the transient dynamics of various materials, for example semiconductors. These studies in turn provided new materials for lasers that could produce even shorter pulse durations with a continuing legacy that has led to the creation attosecond pulses and the shortest ever man-made events. It is precisely because these laser pulses are indeed the shortest events we can make in experiments that most ideas aimed at studying ultrashort laser pulses rely in one way or another to probing, gating or interfering with another equally or shorter laser pulse, thus using light to probe light. Indeed, there is no hope of building a mechanical or electronic shutter that can open and close a camera on the femtosecond timescales required to capture light-in-flight. Yet, various forms of stroboscopic imaging can be adapted to the purpose.

We separate the different capture techniques into three groups: \emph{intensity gating} operates on a similar principle to a conventional camera shutter that transmits only those photons that strike it at a specific time, \emph{coherent or holographic gating} selects photons at a specific time by interfering them with a timed reference pulse, and \emph{continuous capture} methods detect (ideally) all arriving photons together along with a recording of time of their arrival.\\

\subsection{Intensity gating.}
Intensity gated imaging systems consist of a slow imaging sensor combined with a fast shutter or gate that only transmits light at specific times. The light reaching the sensor represents a time slice or frame of the scene time response. The complete time response is recorded by repeatedly illuminating the scene and changing the delay between the illumination pulse and the opening of the gate in each frame. This method can yield very high time resolutions but since it relies on rejecting photons, it tends to require very bright illumination. \\
Mechanical shutters, used also by Muybridge, can only achieve rather limited temporal resolution and are many orders of magnitude too slow to achieve LiF capability. Light, used for example as a stroboscopic illumination or in a single flash, can of course provide the resolution required to freeze motion at the time scales relevant to light propagation. However, light cannot be used to trivially freeze the motion of another light beam. This is simply because, whereas the stroboscopic light will interact with a moving object (made of matter) and scatter back to the camera, it will not interact with another light beam: any other light beams or pulses will be invisible to the stroboscopic illumination. The solution, `nonlinear optical gating', is to use materials that exhibit an optical nonlinearity. For example, if the stroboscopic illumination can modify the medium as it passes through it, this modification will then in turn act on the light that we want to measure in flight and ideally reflect or scatter some of this towards the camera. \\
One may also use electronically gated cameras (`gated image intensifiers'), for example intensified CCD cameras or streak cameras can achieve the temporal resolutions required for LiF.

\subsubsection{Nonlinear optical gating.}
Nonlinear optical gating was probably the first form of LiF imaging. The first experiments were carried out as early as 1967  at Bell Labs who directly imaged 1 ps long pulses propagating in fluorescent dye \cite{1TPF1967}. The key feature in this measurement that is shared in one form or another with all optical gating techniques, is that the pulse, that is both ultrashort and moving at the speed of light, can only be frozen by overlapping it, i.e. gating it, with another ultrashort pulse. In this specific case, the pulse train from the laser was folded back on itself with a mirror so that a reflected pulse would overlap with the successive, counter-propagating pulse. The standing wave pattern would therefore be localised to the overlap region and effectively `freeze' the motion of the light pulse that can now be observed as a result of the higher standing wave intensity and consequent two-photon emission from from the fluorescent dye.\\
An alternative approach was demonstrated shortly later, by another group at Bell Labs  \cite{1969gate,2HSP1971,3HSP1971}. Now the pulse, obtained by taking the second harmonic at 530 nm of an intense pump pulse at 1060 nm,  is allowed to propagate in ``milky water'' (so as to scatter a sufficient amount of light towards the camera): the gating is obtained by using part of the pump pulse at 1060 nm to excite a cell filled with CS$_2$ and placed in front of the camera. CS$_2$ is a highly nonlinear liquid which reacts in 10 ps time scales to intense laser illumination and, among other effects exhibits a transient birefringence. By placing the cell between two crossed polarisers, light at 530 nm is transmitted only in the presence of the pump pulse that therefore acts as an ultrafast, 10 ps gate. The authors use this ultrafast gate to image a light pulse propagating through the milky water, to image satellite pulses in their laser beam and to observe relativistic distortion effects. We shall comment further on the latter effects in a dedicated section below. \\
More recent methods of nonlinear optical gating rely on the subsequent development of second order nonlinearity crystals that are now available in high quality and for a range of compositions and crystal types. Common examples are Deta-Barium-Borate (BBO), Lithium Niobate (LiNbO$_3$) or Potassium Di-hydrogen Phosphate (KDP)  all generate second harmonic or sum-frequency components when pumped by intense laser pulses. By spatially resolving the second harmonic signal generated at the overlap of a probe pulse (that is being imaged) and a ``gate'' pulse, one can recover an image of an object placed in the probe beam path or even recover the shape of the probe light pulse itself by varying the relative pump-gate pulse delay and recording an image of the second harmonic signal at each delay \cite{SH1,SH2}. A schematic overview of this approach is shown in Fig.~\ref{pump-probe}. This optical gating technique has been used for example to study the  evolution of ultrashort laser pulses that undergo severe reshaping due to intense light-matter interactions. By using a gate pulse that is much shorter than the probe pulse being imaged, temporal resolutions of 10 fs or less can be achieved with spatial resolution of the order of a few microns, thus providing a unique insight into the complicated evolution dynamics of femtosecond pulses \cite{SH3,SH4,SH5}. In Fig.~\ref{bullet} we show an example of a LiF tomographic reconstruction of a light bullet propagating in glass: note the longitudinal 200 fs dimension of each frame, corresponding to about 60 $\mu$m and with a  resolution of about 6 $\mu$m in both longitudinal and transverse dimensions. A snapshot of the light bullet is taken every 1.8 mm therefore creating a video that has 8 frames, each separated by $\sim6$ ps, i.e. at more than 0.1 trillion frames/second. This resolution is sufficient to capture the light bullet as it collapses in on itself (frames 1-3) and then slowly develops shock-like patterns around it. Full details of the physics of this process can be found in Ref.~\cite{SH5}.\\
If this LiF pulse tomography is then combined with spectral measurements, it is possible to apply retrieval techniques that allow to even retrieve the full amplitude and phase information of a propagating light pulse. One example relies on capturing the full 3D intensity profile using the time-gating method described above and combining this with an angle-resolved measurement of the spectrum. These two measurements in (r,t) and (k,$\omega$) space, are Fourier-space couples such that applying a Gerchberg-Saxton routine allows to retrieve full amplitude and phase profile of the pulse for various propagation distances \cite{SH6, SH7}. Alternatively, Frequency-Resolved-Optical-Gating (FROG) \cite{FROG} can be adapted to provide full 3-dimensional phase and amplitude information of propagating pulses. FROG essentially relies on the measurement of the light pulse spectrum at varying temporal delay. This is achieved by gating the probe beam with a gate pulse on a nonlinear crystal and then measuring the spectrum for each pump-gate delay. Full 3D reconstruction of the pulse is obtained by combining the spectral measurement at each spatial point with a spatial phase reference measurement that can be obtained using a Shack-Hartmann sensor or through an additional reference measurement \cite{SH9,SH8,SH10}. These measurement techniques have been used for example to reveal subtle features such as the evolution of the superluminal motion of the central peak of femtosecond Bessel pulses \cite{SH8,SH10}.\\

\subsubsection{Femtochemistry and imaging of molecule dynamics.}
Recent advances in imaging the ultrafast dynamics of molecules have built upon pump-probe femtochemistry approaches, first introduced by Zewail and co-workers and provide picosecond or femtosecond temporal resolution \cite{Zewail}. This field would require a whole review on it's own, so we briefly summarise only some of the main recent results (a brief overview can be found in Ref.~\cite{nature_femtochemistry}). Two different approaches have been followed: adding time-resolution to existing spatial imaging techniques and adding spatial imaging technology to existing (or modified) temporal-dynamics measurement methods.\\
For example, Cocker et al. combined scanning tunnelling microscopy with a THz probe beam that uses the oscillating carrier wave of the THz pulses to directly manipulate electronic motion on the light field timescales. The THz field excites a targeted orbital transition in the molecule and allows to record the subsequent  coherent molecular vibrations  directly in the time domain with 100 fs temporal and  0.01 nm spatial resolution \cite{huber}. \\
Another example recently proposed by Wolter et al. uses laser-induced electron-diffraction (LIED) to achieve sub-fs, time-resolved images of the dynamics of bond-breaking in a molecule \cite{Wolter}. LIED effectively extends standard electron diffraction techniques and provides the required temporal resolution by using an electron that is freed from the same molecule being imaged using a control laser pulse, in essence, ``self-imaging of a molecule by its own electrons'' \cite{self}. \\
These are just a few examples that have been mentioned solely to outline that beyond the ``macroscopic'' imaging techniques and applications that we focus attention on in this review,  there is also a very active research community working in the ``microscopic'' domain.

\subsubsection{Ultrafast time-stretch photography techniques.}
\begin{figure*}[t]
\begin{center}
\includegraphics[width=12cm]{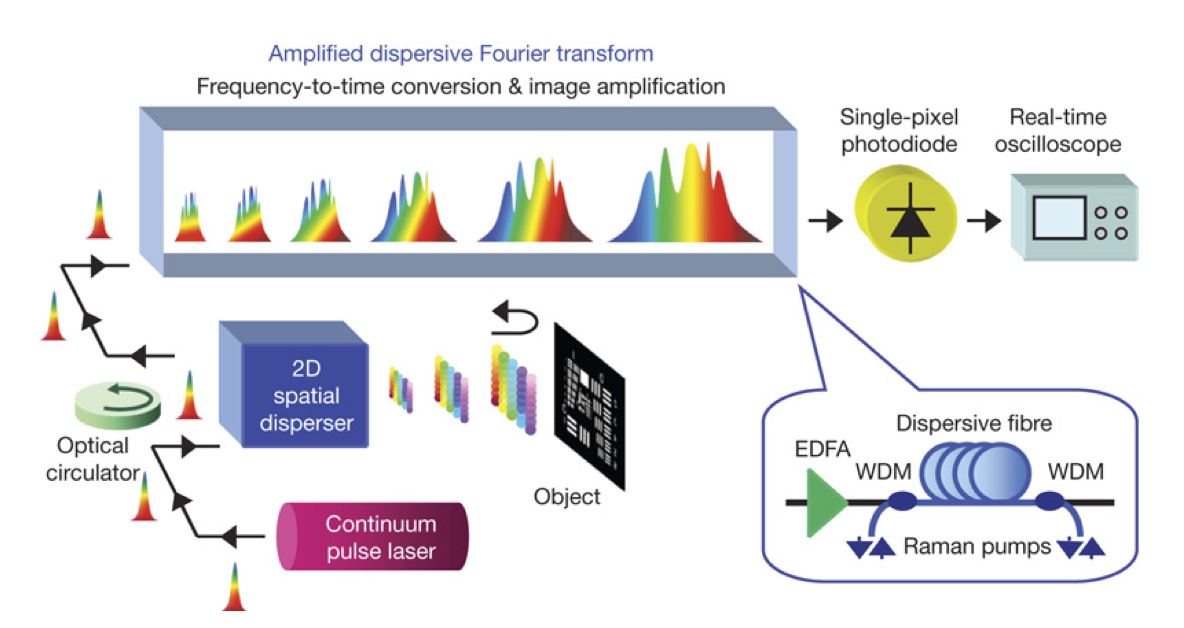}
\caption{Working principle of STEAM. Figure adapted from Ref.~\cite{steam}.}
\label{steam}
\end{center}
\end{figure*}

Time stretch photography relies on encoding different pixels of an image signal in the signal spectrum, followed by optical time-stretching (see Ref.~\cite{time_stretch} for a review). The basic principle of optical time-stretching was introduced by Han et al. \cite{timestretch} and used, for example, to provide the first direct experimental study of rogue wave dynamics by enabling the real-time acquisition of complex optical spectra at 20 MHz repetition rates \cite{rogues}. The basic underlying principle relies on the time-stretching of a broadband optical pulse when this propagates through a dispersive material (different wavelengths propagate at different speeds). A long optical fibre for example can spread a broadband pulse from femtosecond durations up to several nanoseconds: due to dispersion, wavelength information is now encoded in arrival-time and can be read-off with a fast photodiode. This approach was later applied to imaging along a single dimension by Goda et al. \cite{goda}. A broadband pulse is spread out by a grating so as to create an elliptically shaped beam with increasing wavelengths along the major axis that is projected onto a sample (a barcode). The reflectivity profile of the barcode is therefore encoded into the spectrum of the reflected light, which is recompressed to a short pulse by retro-reflecting onto the same input grating. This pulse is then sent through a fibre-based optical time-stretcher and is detected with a photodiode: reflectivity variations along the barcode are transferred into arrival time at the photodiode.
The same authors later extended this technique to full 2D imaging and also demonstrate temporal resolution, with a continuous acquisition at 6.1 MHz and 440 ps time window \cite{steam}. The temporal resolution is obtained by using a pulsed laser source: the separation between one pulse and the next determines the frame rate of the camera and the temporal response of the photodiode determines the time window. A schematic overview of this technique, called `serial time-encoded amplified imaging' (STEAM), is shown in Fig.~\ref{steam}. A notable feature of STEAM is the use of dispersive fibre for time-stretching that can also be used to amplify the signal, therefore overcoming losses in the system and allowing the use of relatively standard single pixel detectors. Recent work has employed single photon avalanche diode (SPAD) arrays as an alternative approach to detecting very weak signals without resorting to signal amplification, albeit at the expense of losing the single shot capability of STEAM \cite{Thomson_fibre}.
It is also worth noting that  in STEAM, since  spatial information of the scene is encoded in the light spectrum, knowledge of the spectral characteristics of the scene is required and the spectral reflectivity of the scene should not change over time.\\
An alternative approach that bears some resemblance to STEAM is called STAMP (Sequentially timed all-optical mapping photography) \cite{stamp}. An ultrashort laser pulse is first split temporally into a series of discrete daughter pulses, each with a different spectral content. These daughter pulses are used to image the target as successive flashes in a stroboscopic image acquisition. The image-encoded daughter pulses are then separated using a device the separates the different wavelength slices into different spatial positions. These are finally directed towards different areas of a CCD camera: each of the individual areas on the camera image  encodes a separate frame of the movie. Therefore, differently from STEAM that achieves temporal resolution by repeating illumination with a number of light pulses (separated and limited by the repetition rate of the laser), STAMP achieves temporal resolution within the duration of a (temporally dispersed) single optical pulse. This allows remarkably high frame rates of 4.4 trillion frames/second with 450x450 pixel spatial resolution in just one single shot. On the downside, the final detection method is limited by the number of pixels on the CCD and thus limits the video to a total of just 5-7 frames. Further developments of STAMP have been applied to imaging of ultrafast (and non repetitive) processes such as laser ablation \cite{stamp2} along with evolutions of the same concept, delivering picosecond temporal resolution across several frames \cite{stamp3}.

\subsubsection{Gated image intensifiers.}
\begin{figure}[t]
\begin{center}
\includegraphics[width=6cm]{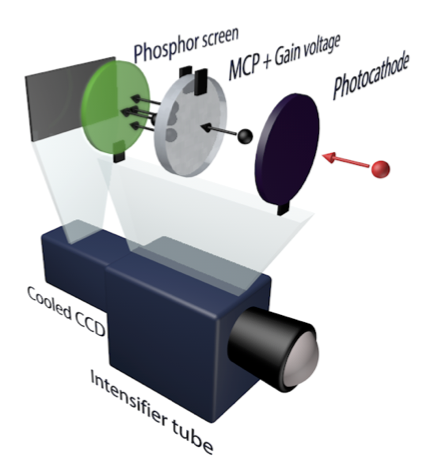}
\caption{Detail of iCCD, showing the individual components. The incoming photon (red sphere) is converted into an electron at the photocathode. This electron is then multiplied in the MCP and then reconverted back into light at rear-end  phosphor screen, where the amplified light can now be detected with a standard CCD. Temporal gating is obtained by controlling the gain voltage on the MCP. \label{iccd_detail}\\}

\includegraphics[width=8cm]{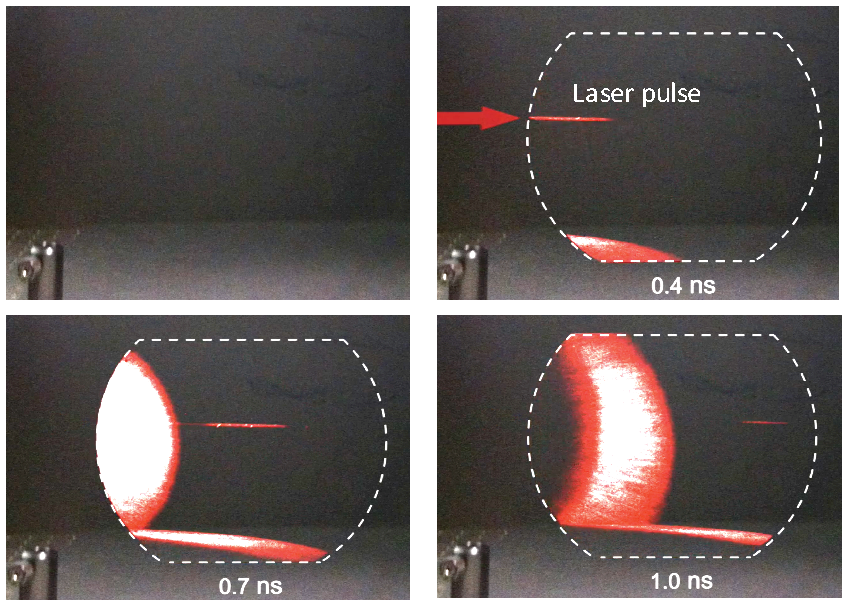}
\caption{Light-in-flight video frames for a laser pulse passing in front of an iCCD camera. Top left: no laser pulse showing the the empty scene; Successive images with indicated acquisition times, i.e. intensifier gate delays (with respect to the trigger signal taken from a reflection of the laser pulse) on the iCCD. The main laser pulse can be seen as a streak, followed by a spherical wave (seen as ellipsoid shapes as it intersects the background) that arises due to a weak diffusion from the inout focusing lens (not visible in the images).}
\label{iccd}
\end{center}
\end{figure}

Image intensifiers are amplifiers that can be switched on and off quickly to function as a gate. The most common instrument for such gated imaging is  the intensified-CCD (iCCD) camera. Intensified-CCDs employ a photocathode inside a vacuum environment followed by a micro-channel plate (MCP). An incoming image signal is converted to an electronic signal at the photocathode. Excited image electrons are accelerated by a strong potential gradient through the micro-channel plate, exciting secondary electrons and amplifying the signal. At the exit of the MCP, electrons strike a phosphor screen that converts them back to photons. By modulating the accelerating voltage across the MCP, the system can be gated with MHz to GHz frequencies allowing light-in-flight videos with time resolutions of a few hundred picoseconds. A schematic overview of this process is shown in Fig.~\ref{iccd_detail}. \\
 Various iCCD models exist with gate times that are typically between a few and several nanoseconds but gates down to 100 ps are also available. The image intensifier produces a secondary effect aside from short time gating, namely sensitivity to extremely low levels of light, all the way down to the single photon level. Indeed, the iCCD has become a widespread instrument in research groups working on quantum imaging and has been used for various applications such as the the direct capture of double-slit interference at the single photon level, imaging of Einstein-Podolsky-Rosen-like correlations, imaging with less than one photon per pixel and real-time imaging of quantum entanglement \cite{miles3,Aspden,miles,miles2,miles4}. In all these applications, the iCCD intensifier is triggered by the second photon of a quantum entangled pair that is generated in a nonlinear crystal and detected on a single photon counter. So this form of quantum imaging also falls under the umbrella of light-in-flight imaging in the sense that the detector is triggered at the correct instant so as to capture the position and arrival time of a photon.\\
In Fig.~\ref{iccd} we show an example of data taken with an iCCD (LAVision GmBH) that has a temporal gate width of 200 ps. The gate is triggered by a reflection from the main laser pulse as this bounces off a mirror. The laser pulse then propagates in free space, from left to right whilst a spherical wave, created by a diffusive reflection from the input focusing lens is also seen as this intersects the background scene. The propagating pulse is seen thanks to the iCCD that is sufficiently sensitive to capture the weak Rayleigh scattered photons from air molecules. The successive images are obtained by retarding the intensifier gate with respect to the input trigger signal by the times indicated in each frame. The laser pulse, generated by a 0.1 W Ti:Sapph oscillator, has a duration of only 100 fs but here is seen to be roughly 10 cm long due to blurring from the 200 ps duration of the iCCD temporal gate.

\subsection{Coherence gating.}
Rather than using an intensity gate analogous to a shutter, one can use the coherence between a reference light source and the received light to accurately and inexpensively determine its time of flight. \\
\begin{figure}[t]
\includegraphics[width=8cm]{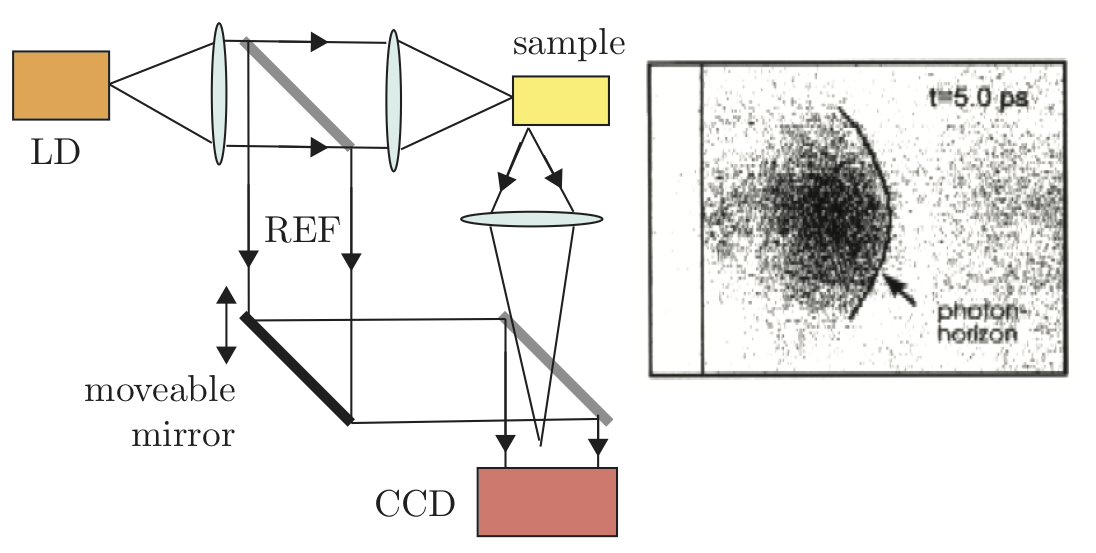}
\begin{center}
\caption{An imaging white light interferometer, using an incoherent light source (LD) that extracts the variations on the signal due to interference fringes as the reference arm mirror (REF) is moved. The scattering sample is observed from the side and the input light is focused closed to the observation facet so as to visualise light as it diffuses through sample. An example of a measurement is shown in the inset where the ``photon horizon'' can be clearly seen (adapted from Ref.~\cite{Hausler:96}).}
\label{white_light}
\end{center}
\end{figure}

\subsubsection{White light interferometry.}
In white light interferometry, coherent gating of the received light signal is obtained by relying on the low coherence length of light illuminating the scene. The returned light is interfered with a reference signal from the same light source: a schematic of the setup is shown in Fig.~\ref{white_light}. When the phase of the reference light is shifted, for example by changing the length of the reference arm, interference fringes are seen by the detector but, due to the short coherence length, interference is observed only over a very short interval effectively gating out only the light that has travelled the same optical path as the reference beam. Coherence lengths as short as 30 $\mu$m are readily achievable thus delivering 100 fs temporal resolution over total observation times of the order of several picoseconds. White light interferometers have been realised with arrayed detectors to provide LiF videos of light propagating in strongly diffusive materials including human tissue and show various physical effects such as the formation of a ``photon horizon'' (a sharp leading edge front in the diffused light beam) and the propagation of diffuse light around absorbing obstacles \cite{Hausler:96}. \\
Gkioulekas et al. later introduced an OCT-inspired interferometric layout that was combined with a computational imaging approach to decompose light transport based on spatial layout, optical path-length and polarization. The system is able to resolve path lengths of just 10 $\mu$m and can be applied to scattering in diffusive media and the characterisation of bulk objects \cite{OCT2}.
\begin{figure*}[t]
\includegraphics[width=16cm]{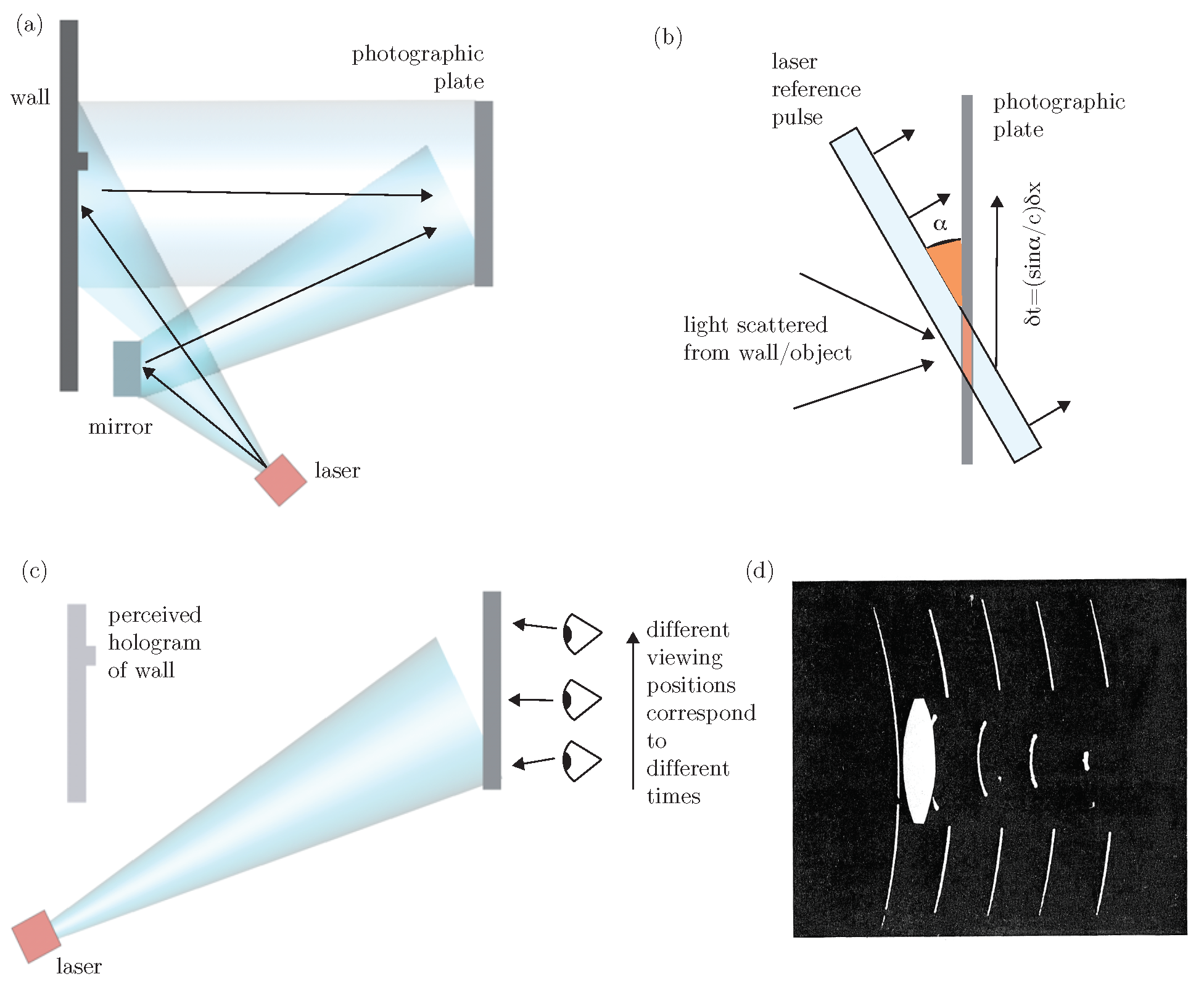}
\begin{center}
\caption{Holographic Light-inFlight imaging. (a) Schematic layout: a pulse laser source illuminates a screen or wall that scatters light towards the photographic plate. A reference pulse, extracted from the main beam with an angled mirror, overlaps the scattered light at an angle $alpha$ with respect to the photographic plate. (b) The two pulses create holographic interference fringes at a specific point on the plate only for light that has travelled identical optical paths, therefore mapping spatial points $x$ on the photographic plate onto evolution time, $t$. (c) The hologram can then be read out by illuminating the plate with the same laser source, placed at the same angle as the reference beam. By moving the eye or camera to different $x$-positions, different times will be seen of the light pulse propagating through the scene and scattering from the wall. (d) An example of such a measurement showing a composite image of 5 different holograms of a 10 ps light pulse illuminating a lens (shown in white as an overlay (not visible in the original hologram).}
\label{fig:abramson}
\end{center}
\end{figure*}

\subsubsection{Holographic light-in-flight.}
The use of holographic plates to record light-in-flight was first demonstrated by Nils Abramson \cite{5LiF1978,6HR1983,7SPH1989,8HTR1991,abramson}. The experimental setup is similar to that used for recording a conventional hologram except that both the object and reference beam are coherent short pulses. A hologram of the scene is formed only where the object and reference overlap on the holographic film. For an appropriately chosen tilt of the reference pulse front, different regions of the holographic film encode different times or scene depths. 
This type of holography is in essence a time-gated viewing technique where the time-gate is provided by the fact that only those parts of the reference pulse and object pulse that have travelled the same optical distance will interfere with each other and thus produce interference fringes (i.e. holographic information) on the photographic plate. Figure~\ref{fig:abramson} illustrates  the basic experimental arrangement. A pulsed laser source is directed onto the scene that one wishes to visualise as light passes across it. In Abramson's original experiments, this consisted of a mirror or a lens placed against a wall such that light from the laser (and then reflected from the mirror or focused by the lens) can scatter from the wall towards the photographic plate. This scattered light is interfered with a reference pulse, split from the main beam with a mirror and directed onto the photographic at an angle, $\alpha$. This angle, in combination with the short pulse illumination, is the key to LiF holography. The illumination geometry on the photographic plate is shown in more detail in Fig.~\ref{fig:abramson}(b): At a given moment in time, the angled wave front intersects the photographic only in a small region (shaded in light red). This intersection point moves across the plate with speed $v=c/\sin\alpha$, i.e. different points $x$ are crossed at different times,  thus effectively mapping spatial positions $x$ onto time $t$. At each position $x$, only light that has travelled the same time $t$ will interfere with the reference pulse and provide a holographic pattern. This pattern can then be read out by illuminating the plate with the same laser and from the same angle as during the measurements, as shown in Fig.~\ref{fig:abramson}(c). By then moving the eye across the plate, a holographic reconstruction of the scene will be observed for different times.
An example of such a hologram is shown in Fig.~\ref{fig:abramson}(d), adapted from \cite{6HR1983}: light is propagating from left to right and the central region of the wavefront intersects, and is then focused by a lens. The figure is the composed image of five separate photographs taken by translating a camera across the photographic plate at positions that differ by about 200 ps.\\
It is worth noting that one does not necessarily need to use pulsed laser sources and indeed the first experiments where performed with a low coherence time, CW laser \cite{5LiF1978}. Moreover, the temporal resolution $\delta t=\delta x\sin\alpha/c$, can be improved by increasing the angle, whereas the total temporal window can be increased by increasing the size of the photographic plate and reference beam.\\
This method of LiF imaging can readily achieve picosecond temporal resolution and is probably to date, still the most precise available method that does not require resorting to nonlinear optical methods described above. It has recently also been implemented using more modern approaches with the photographic plate replaced by a CCD camera \cite{digitalLiF} allowing temporal frames separated by 175 fs to be recorded. \\

\subsection{Continuous Capture.}
While gated techniques rely on the rejection of photons (that lie outside the temporal gate) to achieve precise temporal resolution, there are other devices that seek to capture the entire optical signal as a function of time. This is fundamentally more light efficient, even though the methods described here still do not succeed at capturing and utilising every photon that reaches the imaging system. 

\begin{figure}[t]
\includegraphics[width=8cm]{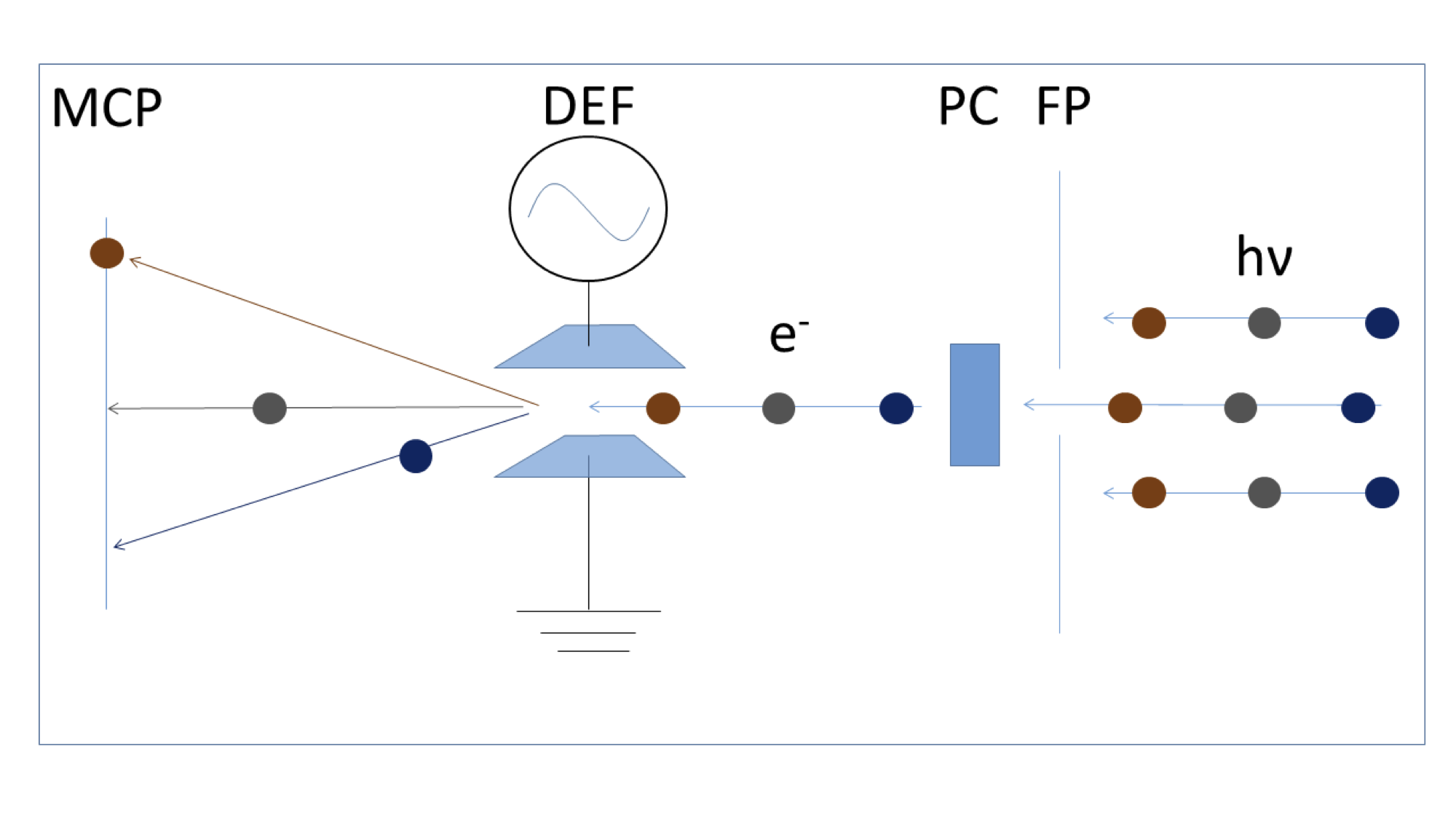}
\begin{center}
\caption{Schematic of a streak tube. Photons from the scene are imaged onto a slit in the objective focal plane (FP). The slit is then imaged onto a photocathode (PC) where photons are converted to electrons traveling electrons are directed by a time varying deflection field (DEF) and are imaged onto a micro-channel plate (MCP) at the end of the streak tube. Figure adapted from~\cite{velten_femto-photography:_2013}.}
\label{streak_cam}
\end{center}
\end{figure}

\subsubsection{Streak Cameras.}
Like image intensifiers, streak cameras use a photocathode to convert the photons in the optical signal to electrons that are then amplified by a micro channel plate (MCP) and converted back to photons by a phosphor screen. In addition they apply a deflecting electric field on the electrons to achieve time resolutions down to hundreds of femtoseconds. A schematic of a streak camera tube is shown in Figure~\ref{streak_cam}. A streak camera images only one line of the scene at a time. Incoming light is first imaged by an objective lens onto a slit that transmits only that one line. After the slit the line is imaged on a photocathode. The excited photoelectrons are deflected by a time varying electric field and imaged onto an MCP where they are amplified. At the exit of the MCP a phosphor screen converts the electrons back to photons that are imaged by a CCD or sCMOS camera. The achievable time resolution in this system is limited by the speed of the deflection signal and the resolution of the electron optics that image the streak onto the MCP. This resolution can be improved with more sophisticated electron optics and higher acceleration of the electron beam. The fastest commercially available streak camera systems are in the range of picoseconds to hundreds of femtoseconds (e.g. Hamamatsu FESCA-100 and C10910, Optronis Optoscope, Sydor ROSS 1000) but much faster systems down to attoseconds have been demonstrated as proof of principle experiments~\cite{PhysRevLett.88.173903}. Besides the complex design, the main drawback of streak cameras for imaging is their limitation to imaging one line at a time. This problem has been addressed in many different ways. If the signal is repeatable, as is the case for a light-in-flight signal created with an active light source, simply scanning the imaging system provides a straight forward solution that yields light-in-flight videos with very high contrast and resolution~\cite{velten_femto-photography:_2013}. Alternative approaches have aimed at optically mapping the streak camera line to a rectangular field of view using pinhole arrays \cite{Landen}, lenslet arrays, fiber bundles, and modified cathode designs~\cite{streak2}. These approaches are able to image events even for single exposures, but yield very low resolution on the order of 20 by 20 pixels. 
 Heshmat et al. proposed the use of a tilted lenslet array \cite{Heshmat14}. Each lenslet in the array projects an image at a different height onto the slit aperture of the streak camera, and both horizontal and vertical information is recorded in a single, multiplexed streak image with 2 ps temporal resolution.
The same year, Gao et al. proposed Compressed Ultrafast Photography (CUP). This technique using a compressive sensing approach to transform a streak camera into a single shot, full 2D imaging device that can capture events with up to $10^{11}$ frames per second \cite{CUP1}.
The spatial information is encoded with a pseudo-random binary pattern that is imparted with a digital mirror device (an array of tunable micro-mirrors), followed by a shearing operation in the temporal domain that is performed by using a streak camera with a fully opened entrance slit. This
encoded, sheared three-dimensional (x,y,t) information is recorded as a single snapshot on  the back-end CCD camera. The image is finally reconstructed
by iteratively estimating a solution that minimizes a cost function that measures the difference between the measured sheared image and the simulated sheared image for a given initial guess solution (se e.g. \cite{compressive}).\\
This computational imaging approach was further improved by adding an additional prior to the cost-function minimisation routine in the form of an un-sheared image of the scene \cite{CUP2}. This method, tested for example on reconstructing images of pulses propagating in air, shows improved definition of the final results with respect to the original CUP method.
\begin{figure}[t]
\begin{center}
\includegraphics[width=5cm]{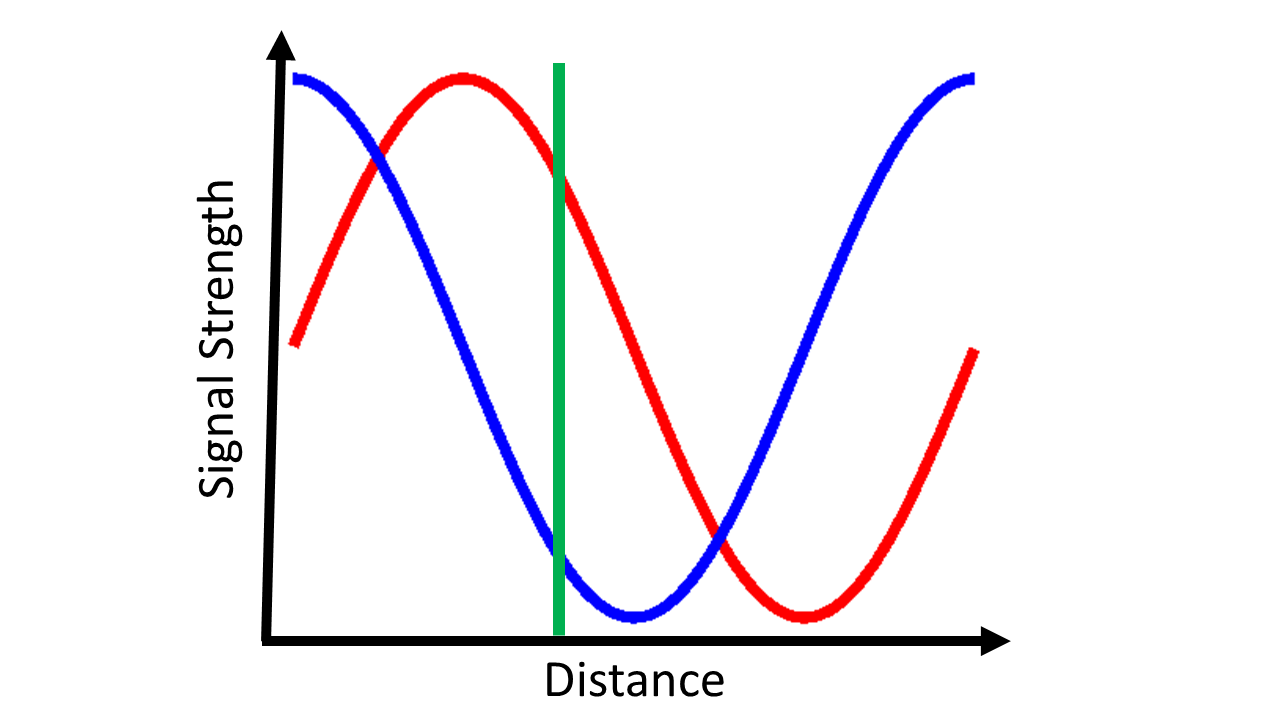}
\caption{The in phase and in quadrature signals captured by a PMD pixel observing a surface as function of surface distance or light travel time. From a single measurement pair of in phase and in quadrature signals the distance can be determined by finding the corresponding value pair on this plot. For example for the value pair highlighted by the green line.}
\label{fig_pmd}
\end{center}
\end{figure}
\begin{figure*}[t]
\begin{center}
\includegraphics[width=14cm]{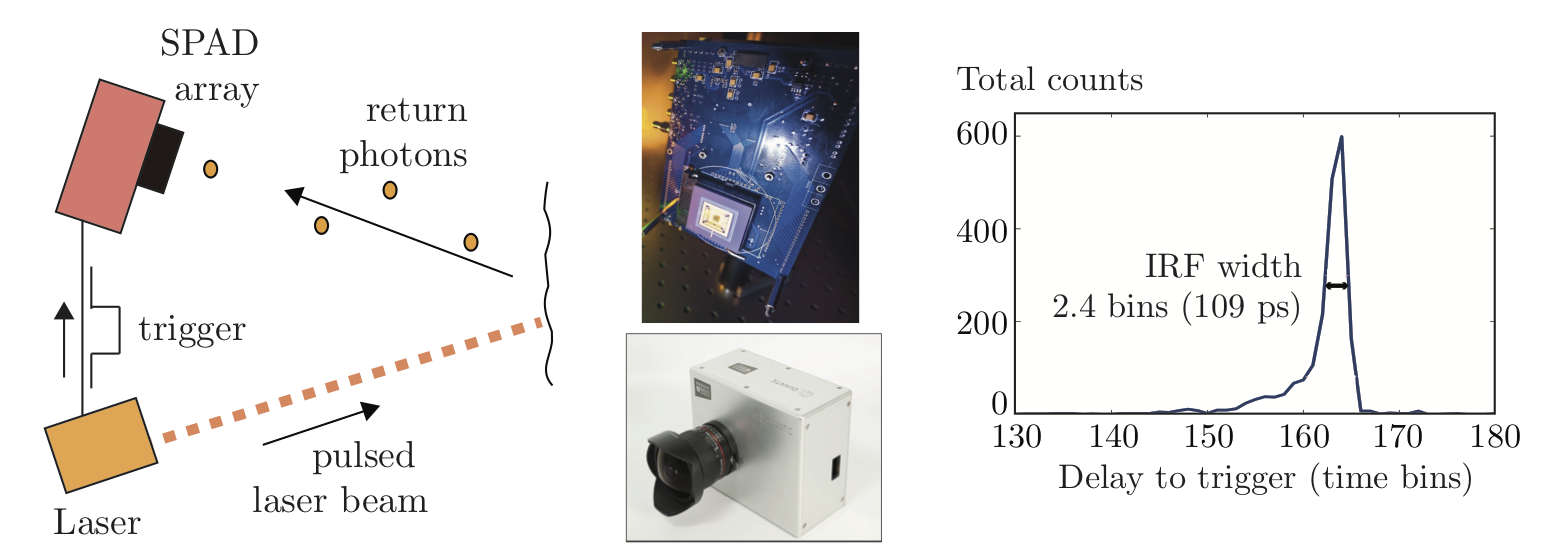}
\caption{Left: schematic of TCSPC with active illumination of a scene with a pulsed laser. An electronic signal from the laser, synchronised to the laser pulses, triggers the SPAD detector or array. The SPAD then detects return photons from the scene and assigns to each photon, a time bin according to the measured arrival time with respect to the laser trigger. Middle: photos of a SPAD array electronics board and the packaged device by QuantIC \cite{quantic}. Left: an example of a measured IRF histogram for just one pixel of the MF32 array, using 100 fs pulses reflected from a flat surface.}
\label{spads}
\end{center}
\end{figure*}
\subsubsection{Photonic Mixer Devices.}
Photonic Mixer Devices (PMD) are CMOS sensors that are commercially available at low cost for time of flight ranging applications such as the microsoft kinect game controller~\cite{Langmann2012DepthCT}. The same technology is also marketed under the names CanestaVision and Swiss Ranger. In these devices the scene is illuminated with a time varying intensity of the form $I(t)=I_0\cdotp \sin(\Omega t)$ where $\Omega$ typically varies from 20 MHz to about 100 MHz. The PMD sensor then determines the phase shift of this sinusoid that is the result of propagation in the scene. Each pixel collects two images with modulated gain in each pixel $S_P=\int{I(t+\delta t)\cdotp\sin(\Omega t)dt}$ and $S_Q=\int{I(t+\delta t)\cdotp\sin(\Omega t+\pi)dt}$. From these two images the time delay $\delta t$ due to light propagation in the scene can be computed. This is illustrated in Figure~\ref{fig_pmd}. The Figure shows a plot of the in phase and in quadrature values collected as a function of travel distance. An individual measurement then just has to be matched against this plot to determine depth.\\

In the absence of background light two phase shifts are sufficient. In most cases background light can not be ignored and the signals include a constant background. In this case at least three measurements with different phase-shifts are required. Since a PMD captures two phase shifts in one image, a PMD measurement typically uses two PMD images and four phase shifts of 0, $\frac{\pi}{2}$, $\pi$, and $\frac{3}{2}\pi$. From these 4 measurements the depth can be recovered despite the unknown scene reflectance and the unknown constant background illumination.\\

PMDs are among the most light efficient high speed capture methods today. Due to large scale adaptation in ranging applications the sensors are inexpensive and technologically mature. However, current PMD designs do not work at modulation frequencies above 100 Hz~\cite{Luan01}. The technological maturity and specialization for specific applications is also a disadvantage of PMDs for experimenters. Commercially available models typically do not allow direct access to the modulation voltage signal and therefore make it hard to apply the devices to anything but ranging applications. But maybe more importantly, the use of standard photodiode technology for the collection of light implies a relatively limited sensitivity when compared for example to intensified sensors or single-photon sensitive devices described below. This limited sensitivity hinders applications to scenes that extend beyond a few meters from the detector.
Nonetheless, PMDs have been used successfully for light-in-flight imaging \cite{Heide1,Raskar_PMD} and Non-Line-of-Sight imaging \cite{Heide2} applications.

In these applications the PMD concept typically has to be extended to multiple PMD frequencies to allow the simultaneous detection of multiple phase shifts that are due to multiple reflections from partially translucent surfaces.

\subsubsection{Single Photon Avalanche Diodes.}
Single Photon Avalanche Diodes (SPADs) are semiconductor diodes that are reverse-biased at a voltage close to the breakdown voltage. When absorbing a photon, the resulting photoelectron is accelerated and excites secondary electrons creating a self-sustaining avalanche. The resulting current is detected and a quenching circuit lowers the bias voltage across the diode to stop the avalanche and flush all excited carriers from the diode junction. During this quenching process the diode is not sensitive to photons. After a fixed dead time, ranging from hundreds of nanoseconds to several microseconds depending on the SPAD type, the bias voltage is again raised to just below the breakdown voltage and the SPAD is ready to receive the next photon. The output of the SPAD is an electronic impulse for each photon that is absorbed in the diode outside of the dead time.\\
In order to create a time profile for a light-in-flight video, the timing of SPAD signals has to be correlated with a clock signal and compiled into a temporal histogram. This process is referred to as Time correlated Single Photon Counting (TCSPC).

\paragraph{Time correlated single photon counting} (TCSPC) is a technique that provides sensitive detection of signals at high time resolution through the use of photon counting sensors \cite{TCSPC}. TCSPC is typically used in areas such as fluorescence lifetime measurements and quantum optics and was first implemented with photomultiplier tubes (PMT) and only more recently with SPADs. Fluorescence lifetime measurement, similarly to LiF, requires the measurement of transient signals with 1-100 ps sampling resolution in order to precisely characterise the fluorescence signal decay rates. Such temporal resolutions are very hard to achieve with standard photodiodes and oscilloscopes. Moreover, if the signal is extremely weak, the discrete nature of the emission that will be composed of maybe just a few photons or less, will also prohibit analog sampling. TCSPC provides an answer to these challenges. The TCSPC electronics are ``initialised'' or triggered by an electronic signal that is provided either by the pulsed laser source (detected with a photodiode) used to illuminate the fluorescent molecule or LiF scene or, in the case of a quantum optics experiment,  can  also be provided by one photon of a correlated pair (that is detected with a dedicated SPAD). The arrival of a photon, detected on the PMT or SPAD will then create a second electronic signal. With sufficiently precise electronics, the temporal delay between the trigger and the measurement can be read off with high temporal resolution. Each photon count is then time-binned: for example if a 50 ns time window is digitised with a 10-bit time-to-digital converter (TDC), then the arrival time of a single photon count will be inserted in a histogram with $\sim50$ ps width. These concepts are summarised in Fig.~\ref{spads}. \\
\begin{table*}[t!]
\centering
\begin{tabular}{|l|c|c|l|c|c|c|}
\hline

\textbf{\begin{tabular}[c]{@{}l@{}}\\Method\\ $\,$ \end{tabular}}   & \textbf{year}             & \textbf{Ref.}    & \multicolumn{1}{c|}{\textbf{limitations}}                                                                                        & \textbf{\begin{tabular}[c]{@{}c@{}}temporal \\ resolution\end{tabular}} & \textbf{\begin{tabular}[c]{@{}c@{}}single\\  shot\end{tabular}} & \textbf{\begin{tabular}[c]{@{}c@{}}compact\\ portable\end{tabular}} \\ \hline
\textbf{\begin{tabular}[c]{@{}l@{}}\\ light-in-flight \\ Holography\\ $\,$  \end{tabular}}                                                       & 1978                      & \cite{abramson}                 & requires scattering screen                                                                                                                               & 0.1-100 ps                                                              & Yes                                                             & Yes                                                                 \\ \hline
\textbf{\begin{tabular}[c]{@{}l@{}}\\Streak camera, \\ iCCD\\ $\,$ \end{tabular}}       & 2011                      & \cite{velten}                   & \begin{tabular}[c]{@{}l@{}}1. requires scattering medium\\ 2. requires scanning of scene\end{tabular}                                                    & 1-200 ps                                                               & No                                                              & No                                                                  \\ \hline
\textbf{\begin{tabular}[c]{@{}l@{}}\\CUP\\ $\,$ \end{tabular}}                                                                  & 2014                      & \cite{CUP1}                      & \begin{tabular}[c]{@{}l@{}} post-processing relies\\ on scene sparsity \end{tabular}                                                             & 10 ps                                                                   & Yes                                                             & No                                                                  \\ \hline
\textbf{\begin{tabular}[c]{@{}l@{}}\\Photonic \\ Mixer \\ Device\\ $\,$ \end{tabular}}  & 2013                      & \cite{Heide1,Raskar_PMD}                      & \begin{tabular}[c]{@{}l@{}}1. ambiguous inverse problem \\ 2. low  sensitivity limits  range\end{tabular} & 1 - 10 ns                                                          & No                                                              & Yes                                                                 \\ \hline
\textbf{\begin{tabular}[c]{@{}l@{}}\\ SPAD arrays/\\ single pixels\\ $\,$ \end{tabular}} & \multicolumn{1}{l|}{2015} & \multicolumn{1}{l|}{\cite{me1,7VSPAD2015}} & \begin{tabular}[c]{@{}l@{}}TCSPC requires high \\ rep. rate lasers\end{tabular}                                                                          & 50-300 ps                                                               & No                                                              & Yes                                                                 \\ \hline
\end{tabular}
\caption{Summary of main ultra-high speed photography technologies that have been used for LiF imaging, with their main features and limitations. These and other methods are discussed in more in the main text. ``Year'' indicates the year in which the technique or technology was first used in the context of LiF imaging.}
\label{table}
\end{table*}
Several factors will ultimately limit the temporal resolution and thus determine the overall temporal Impulse Response Function (IRF). A key factor is the design of the PMT or SPAD itself. Considering the case of a SPAD, the thickness and absorption coefficient of the material will determine if the photons are mainly absorbed close to the surface of the SPAD or deeper inside the SPAD, i.e. closer or further from the photo-electron collection regions. Similarly, the wavelength-dependent absorption coefficient will lead to wavelength dependent IRFs. One must then also add to this the electronic jitter. All combined, these effects lead to IRFs ranging from 20 ps for PMT-based systems to 100-300 ps for SPADs.\\
As mentioned above, the discrete nature of single photon detection does not allow, with a single measurement alone, to determine the shape of the photon or laser pulse: the full temporal profile is reconstructed by repeating the measurement many times. Typically, one will aim for a final histogram that is populated with $10^2-10^4$ photon counts. If the measurement is performed in the photon starved regime, i.e. with much fewer photon counts than laser pulses, the probability of receiving a photon within the temporal duration of the overall scene is uniformly distributed. The accumulated measurements will then reproduce the correct time evolution of the scene. By `photon starved' one typically assumes that only 1\% or fewer laser pulses contribute to a photon count, although often 10\% will often still be quoted as `starved'. Therefore, a histogram with $10^4$ counts will require at least $10^6$ illuminations. If the scene is illuminated with an 80 MHz repetition rate laser, such a measurement can in principle be acquired in just over 10 ms and easily accommodate for video frame rate measurements. \\
SPAD systems are commercially available as single pixel systems with separate TCSPC units and are widely used for both fluorescence lifetime and quantum optics measurements. Recent developments have also led to the first SPAD arrays with integrated TCPSC using standard CMOS manufacturing techniques \cite{RH2,RH3,RH4,RH5,Tosi2,Tosi3,Cammi,Tosi4}. Currently, typical arrays have 32x32 pixels but higher 64x64 or different formats with similar total pixel counts are available \cite{RH1,Tosi1,linospad}. These are also starting to become commercially available \cite{PF,MPD}. Whereas single pixel SPADs require a separate TCSPC electronics box, each individual pixel in the SPAD array contains both the sensitive detection area and its own TCSPC electronics, therefore providing an extremely compact device (similar size to a standard CCD camera) that is capable of delivering a full 3D (x,y,t) matrix of data providing a temporal histogram for each spatial coordinate (x,y).  IRFs for currently available SPAD arrays is of order 100-150 ps with a typical example shown in Fig.~\ref{spads}. However, due to the presence of the pixel-by-pixel integrated TCPSC electronics, the sensitive area is only of order of a few percent, thus severely limiting overall light collection efficiency. This can be overcome by vertically stacking the pixels so that the photon sensitive area is isolated and placed vertically on top of the TCPSC electronics. This technology, first developed in InGaAs SPAD arrays \cite{itzler1,itzler2,itzler3} has been developed also for Silicon SPADs \cite{RH-vertical}. Notwithstanding these limitations, which are being addressed and will be overcome in future devices, SPAD detectors and TCSPC have allowed a series of breakthrough demonstrations of LiF photography and are allowing the development real-world applications that will be discussed below.
A key point of this technology is the promise for integration into very compact devices. LG were the first to commercialise a phone handset that incorporates a SPAD array with a laser to assist the camera autofocus at short distances \cite{LG}. The LG phone is performing only simple time-of-flight ranging, but nevertheless points towards future devices where LiF imaging will be available in very compact devices.\\
Table~\ref{table} summarises some of the main features of ultra-high speed photography techniques. The table focuses on the methods that have seen the most widespread application for the specific scope of light-in-flight photography: each approach has its own strengths and weaknesses that will make one or the other more attractive depending on the final application and equipment availability in the lab.

\section{Applications.}
\begin{figure*}[t]
\begin{center}
\includegraphics[width=16cm]{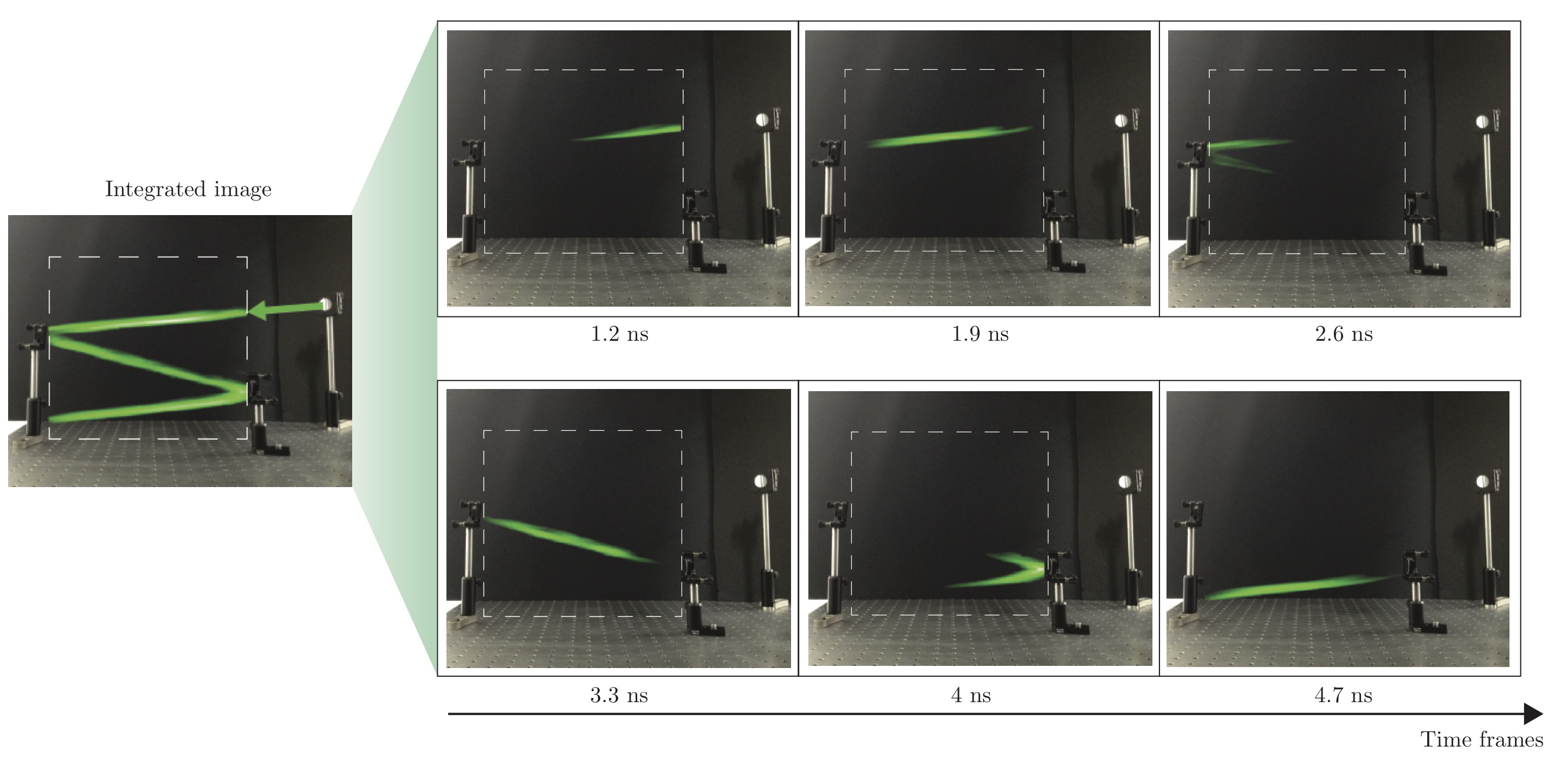}
\caption{Light-in-flightight-in-flight photography of a laser pulse, moving in free space and reflecting from two mirrors. The backdrop, showing the scene and mirrors in color, is a superimposed photograph taken with a cell phone. The dashed box highlights the area imaged by a 32x32 SPAD array. Photons scattered (by air molecules) towards the array from a 532 nm wavelength, 500 ps duration laser pulse are detected and time-binned, achieving a temporal resolution of 120 ps (limited by the IRF of the array) and thus allowing to directly image the propagation of the pulse. Color is applied to the laser pulse in post-processing. More details can be found in Ref.~\cite{me1}.}
\label{lif}
\end{center}
\end{figure*}
\subsection{Light-in-flight photography.}
The most direct application of light-in-flight imaging is the presentation of the video itself for scientific, educational, or artistic purposes. Various authors have applied their techniques to this end starting with Nils Abramson's light-in-flight holography \cite{5LiF1978,abramson}, followed two decades later with white light interferometry \cite{Hausler:96}. Both these systems rely on coherence and provide an image of the coherent electric field amplitude of the light rather than it's intensity. More recently a streak camera based system was used to provide intensity videos of light-in-flight that are among the highest quality light-in-flight videos to date \cite{velten}. SPADs have been used to image laser pulse propagation in air (i.e. without the requirement of any additional scattering medium) and as a way to reduce the footprint of the experimental setup \cite{me1}. An example of such a measurement is shown in Fig.~\ref{lif}, where a 500 ps laser pulse is captured as it propagates, bouncing between two mirrors. \\
One of the main motivations behind LiF photography has been simple curiosity and fundamental questions regarding the nature of light propagation. A key example will be given below, where we describe how relativistic effects connected to the finite propagation speed of light, can be readily observed with LiF photography, unveiling distortions of space-time connected events that had never been observed before. \\
As with all photographic techniques, once the basic technology is mastered and understood, attempts follow to harness the results for artistic purposes. Examples are original videos prepared by N. Abramson \cite{abramson_video}, museum exhibitions featuring LiF videos reproducing the first measurements of the speed of light performed by Foucault \cite{lily1} and LiF portraits \cite{LiF_portrait}. It is no coincidence that one of the pioneering works in the digital age of LiF  photography \cite{10SA2011}, was entitled ``Slow Art with a Trillion Frames Per Second Camera'', suggesting the viewpoint that this is a form of art, with applications for technology and fundamental science.\\

\subsection{Non Line of Sight Imaging.}
\begin{figure}[t]
\includegraphics[width=8cm]{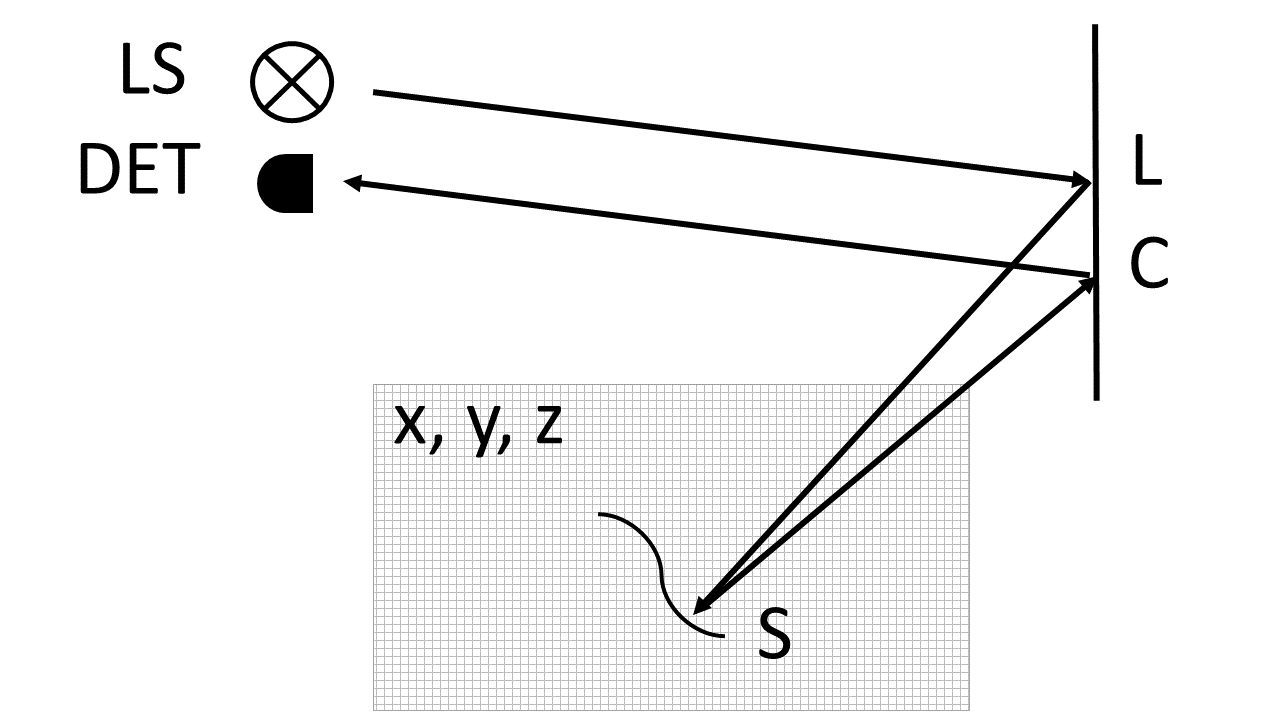}
\begin{center}
\caption{Light path in a Non-Line-of-Sight imaging experiment.}
\label{NLOS_setup}
\end{center}
\end{figure}

Non-Line-of-Sight (NLOS) imaging refers to the ability to image and collect information about a scene that cannot be directly viewed. A typical example is a scene or object that is hidden behind a wall or inside a room. Optical NLOS imaging uses light-in-flight videos of a scene to compute 3D geometry, albedo, and reflectance information of parts of the scene that are not directly visible to the light-in-flight camera. 
A NLOS measurement typically involves illuminating a set of points $L$ on a surfaces in the visible scene sequentially with a short pulse (Figure \ref{NLOS_setup}). Light returning from a set of points $C$ on the scene surface is recorded as a function of time resulting in a set of time responses $D(L,C,t)$ with one time response associated with each pair of illumination ($L$) and detection ($C$) points. The hidden scene geometry is typically represented as a voxel space $V(x,y,z)$ with each voxel representing the albedo of an object surface patch located at a point with coordinates $(x,y,z)$. It is straight forward to compute a dataset $D$ from a known hidden scene $V$ using geometric optics in a process we call forward projection or transient rendering.
\begin{equation}
\mathcal{F}(V)=D
\end{equation}
To reconstruct a hidden scene we need to find the inverse of this operation
\begin{equation}
\mathcal{F}^{-1}(D)=V
\end{equation}
Typically, $\mathcal{F}$ is approximated as a linear operator $A$ and a linear inverse is computed, for example using algebraic reconstruction~\cite{gordon_algebraic_1970}, or convex optimization~\cite{heide_diffuse_2014}. This approach works well in scenes that do not violate the linearity assumption. Scenes containing occlusion, multiple reflections between hidden surfaces, or non-lambertian surfaces where albedo depends strongly on surface angle lead to substantial non-linear contributions in the light transport process $\mathcal{F}$. Nonlinear inverse methods have had limited success to date in reconstructing these scenes.
\begin{figure}[t]
\includegraphics[width=8cm]{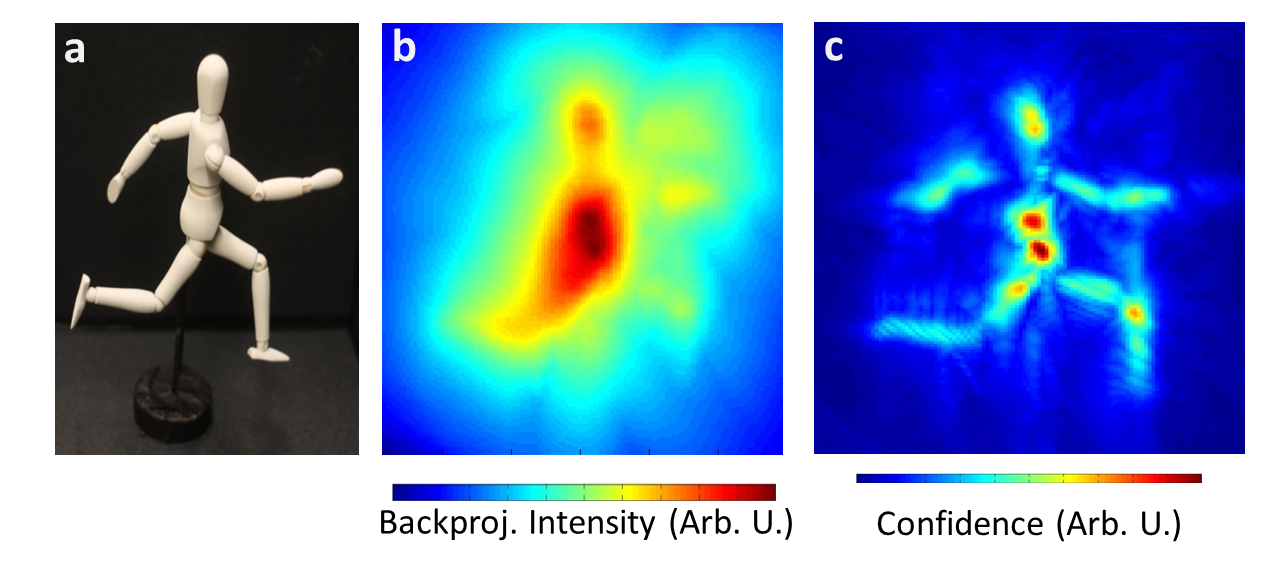}
\begin{center}
\caption{(a) Image of an object imaged. Filtered (b) and unfiltered (c) backprojection of a non-line-of-sight object. Figure adapted from~\cite{velten_recovering_2012}.}
\label{NLOS_result}
\end{center}
\end{figure}
An approach that is robust to deviations from a linear model is filtered backprojection~\cite{velten_recovering_2012}. The backprojection computes an approximate linear inverse followed by a filtering step to extract scene surfaces. Referring to Fig.~\ref{NLOS_setup}, it is assumed that all light collected in D (detector) is from photons having undergone exactly 3 reflections, one at $L$, one at $C$ and one at a hidden object surface point $S$. Every sample in D is projected on all the voxels in V that are possible locations of the surface that reflected the photon by adding the value D to each of the candidate voxels V. Each sample in $D$ thus projects onto an ellipsoid with foci $L$ and $C$ and semi-major axis $a$ that is determined by the travel time between $L$ and $C$ This can be expressed as
\begin{equation}
V(x,y,z)=\sum_{L,C}{D(L,C,c\cdotp d(L-C-(x.y.z))}
\end{equation}
where $d(L-C-(x.y.z))$ is the total round trip distance distance from the light source to the camera via points $L$, $(x,y,z)$, and $C$. An unfiltered backprojection is shown in Figure~\ref{NLOS_result}b. A simple and robust way to extract surfaces from this reconstruction is to apply a laplacian filter to the backprojection result. The result of this operation is shown in Figure~\ref{NLOS_result}c.\\
Time of flight based NLOS imaging was first demonstrated by Velten et. al. using a mode-locked laser and a streak camera with an optimal time resolution of 2 ps. The effective impulse response resolution of the captured data was about 15 ps. The scene was a table-top scene of about 30 centimeter diameter and several objects were reconstructed with a resolution of about 1 cm. Later experiments focused on extraction of other scene properties, such as the  bidirectional reflectance distribution function  (BRDF) \cite{naik_single_2011}, on improvement of the reconstruction algorithm, and on the use of more compact hardware. Heide et. al. demonstrated NLOS imaging using photonic mixer devices (PMD) and presented an improved reconstruction algorithm based on convex optimization. Buttafava et. al.~\cite{buttafava_non-line--sight_2015} demonstrated NLOS imaging using gated SPAD sensors.\\
SPADs have also been applied to the reconstruction of hidden scene with linear arrays \cite{wetzstein1}. Finally, a confocal arrangement of a single pixel SPAD (i.e. the single pixel collection is aligned along the laser beam and both are therefore scanned together, as in standard LIDAR techniques) can lead to a significant simplification of the 3D image retrieval \cite{wetzstein2}. The success of this last approach lies both in the simplicity of the setup that uses only a laser and a single pixel SPAD that are then scanned across the scene together, and also in the very fast 3D retrieval times, approaching real-time, NLOS 3D imaging.

\subsubsection{Gated detection, continuous illumination and deep learning.}

Alternative methods have also been proposed for identifying objects that are hidden from the direct line of sight. \\
Laser gated imaging uses short illumination pulses and the return signal from each pulse is recorded with a camera, typically a fast CMOS sensor, that can be electronically gated (i.e. light is collected only within a short temporal window). The delay, $\tau$, between the laser illumination pulse and the electronic camera gate,  determines the distance, $d=2c\tau$ within the scene that is illuminated and viewed on the camera. By sliding the gate temporal delay, a full 3D reconstruction of the scene is obtained. For NLOS higher temporal resolution is required and this was achieved by employing an iCCD camera in place of a CMOS sensor and allowed sampling with gate separations of 67 ps. A computational technique similar to the inverse Radon transform used in computed tomography (CT), allowed the 3D reconstruction of a 20 cm x 20 cm screen placed 2 m away from the observation wall \cite{Velten_gated}.\\
Hullin et al. demonstrated the ability to track single objects in high contrast scenes using only intensity information from a regular CCD camera \cite{hullin}. Even with continuous laser illumination, the return signal from the hidden object that now has no time-dependence and appears simply as a diffuse ``glow'' originating from behind the obstacle will exhibit subtle variations depending on the shape and reflectivity of the hidden scene. A forward modelling, light field propagation approach is used whereby initial guess objects are divided into voxels. The return signal from each voxel is simulated and the voxel shapes/number are then varied in a loop until the difference between the simulated and measured signals is minimised.\\
Caramazza et al., have also demonstrated that is possible to use a deep learning approach to correctly retrieve  both position information and the identity of a person hidden behind an obstacle \cite{pier}. The method relies on the use of a 32x32 SPAD array camera that is used as an array of 1024 independent, {\emph{single}} pixel SPADs. A single acquisition therefore provides 1024 measurements on which to train a supervised neural network: three different individuals are placed in different positions behind a wall and the recorded data is labelled accordingly. This data proved to be sufficient to then correctly identify the position and the name of the individual when tested on new data that had not been used in the training phase. Remarkably, the neural network was capable of correctly identifying the individuals even when these had exactly the same clothing (both in training and testing), this indicating that the temporal histogram from a single pixel contains sufficient information to distinguish subtle changes in the conformation of objects that generally speaking, are rather similar (we were testing only on humans) and differ only in aspects that vary by less than 10\% from one individual to another, such as body height or shape. Further investigation is required in order to evaluate how many individuals can be categorised from information contained in a single histogram (that has only 1024 bits of information) and if it is possible to perform actual imaging tasks with neural-network-assisted reconstruction.

\subsection{Tracking of moving targets around a corner.}
{Acquiring the information necessary to reconstruct a fully detailed 3D image of a scene from behind a corner can be demanding both in terms of hardware and in terms of measurement/processing times. However, there are many NLOS scenarios where a full 3D reconstruction is not required and much lower-level information is sufficient. Examples are the ability to detect motion from behind a corner, locate an (un)known object and/or track it's motion, without actually ever attempting to reconstruct the shape of the hidden object and scene. This is a much simpler problem to solve and can be achieved with less information, at higher speeds and with simplified hardware. \\
\begin{figure}[t]
\begin{center}
\includegraphics[width=8cm]{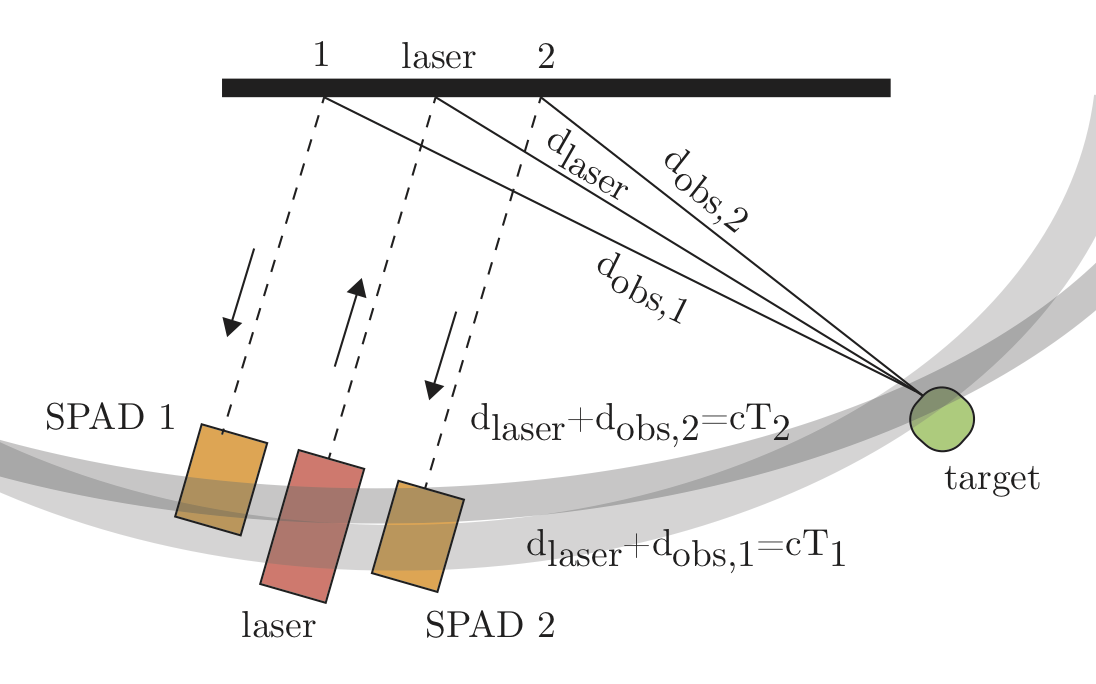}
\caption{Reconstruction of NLOS target position through triangulation: the target is illuminated by a laser, reflected from a diffusive surface or screen. The return light echo, reflected from the target back towards the screen is detected by the SPADs that are operated in TCPSC mode and can therefore determine the time, $T$, for light to travel from the laser spot to the target (distance $d_\textrm{laser}$) and back from the target to the observation spot (distance $d_\textrm{obs}$). The two different detectors will give two separate probability ellipses for the target location (light gray curves). The overlap (multiplication) of these two ellipses (darker shaded, `banana' shaped region) provides an estimate of the target location.}
\label{triangulation}
\end{center}
\end{figure}
%
%
%
Pandharkar et. al. \cite{velten_motion} presented a method to detect changes in a scene from behind a corner and estimate both the size of the moving object and it's direction of motion. Measurements are performed using a pulsed laser in combination with a streak camera and a scanning system to visualise a broad region on an observation surface, i.e. a similar setup to that used for full 3D scene reconstruction \cite{velten}.  By subtracting successive frames, it is possible isolate moving objects from a cluttered background environment. A constrained least-squares problem is then solved in order recover the NLOS target location. Gariepy et. al. presented a method to track multiple objects in real-time (1 Hz refresh rate). This approach relied on the use of a pulsed laser source for the scene illumination and detection with a SPAD camera, thus removing the need for any scanning elements \cite{me2}. A simplified reconstruction method was also implemented that essentially reduces the problem to a triangulation of the hidden target position. Referring to Fig.~\ref{triangulation}, the SPAD detector observes a given point, ``1'' on a screen that intercepts the return light echo from the hidden target, and will measure the time, $T$, for light to travel from the laser spot position to the target (distance $d_\textrm{laser}$)  and back to the observation point (distance $d_\textrm{obs,1}$):\\
\begin{equation}
d_\textrm{laser}+d_\textrm{obs,1}=c\times T_1
\end{equation}
where $c$ is the speed of light. This equation describes the locus of points that are equidistant from the laser and observation spots, i.e. it describes an ellipse that we can treat as a probability distribution for the target position. Clearly, a measurement based on a single observation point is not sufficient to locate the target position. But if we add a second observation point, then we can add a second equation $d_\textrm{laser}+d_\textrm{obs,2}=c\times T_2$. The joint probability distribution is found by multiplying these two individual observation point distributions resulting in a ``banana shaped'' distribution that locates the target position with an uncertainty that is related to the width of the individual ellipses (determined by the impulse response function of the SPAD detector). Increasing the number of detection points, for example by employing a 32x32 SPAD array, will reduce the uncertainty of the final result. Alternatively, the precision can be increased by increasing the detection baseline, i.e. the spatial separation between the observation points. Whilst the SPAD camera requires a standard photographic lens for imaging, which in turns limits the numerical aperture and triangulation baseline, using individual single pixel detectors, each with separate collection optics allows to arbitrarily increase the baseline. 
Under typical circumstances one can make a reasonable assumption regarding the height of the hidden target (if e.g. this a person or a vehicle) so that the measurement of the arrival times of the return laser-echo with only two detectors is sufficient to uniquely identify the position or track the hidden target in the horizontal plane. The additional advantage gained in using single-pixel detectors (aside from off-the-shelf availability) is the ideal fill factor, i.e. ratio of the illumination area on the detector with respect to the sensitive area size on the detector that can be close to 100\% compared to the $\sim1-3$\% of the SPAD camera used in Refs.~\cite{me1,me2}. This enhanced photon collection/detection efficiency allowed to locate and track the motion of human targets from a stand-off distance of more than 50 meters with a 1 Hz acquisition rate \cite{SusanFar}.}\\

\subsubsection{Photon budget.}
{ It is worth noting that in these experiments, laser powers are typically in the 0.1-1 W range. For example, the long range motion-tracking experiments where performed with 500 mW laser power \cite{SusanFar},  corresponding to $\sim10^{18}$ photons/second. Yet only $10^3-10^4$ photons/second are measured in the return signal. This gives a rough feeling for how demanding these measurements can be: the multiple scattering process leads to an effective photon loss of order of $10^{12}$ (best case scenario). This is a strongly photon-starved regime and is the ideal scenario for SPAD detectors. \\
 A related and relevant question is the impact on such single photon sensitive devices of daylight illumination, a necessary consideration when thinking of future applications for e.g. the automotive industry or surveillance. Indeed, the measurements in Ref.~\cite{SusanFar} were performed in a daylight illuminated environment: background sunlight was simply filtered out by placing $\sim1-10$ nm bandwidth interference filters in front of the SPAD detectors so as to isolate  the laser illumination wavelength (1500 nm). A back-of-the-envelope calculation allows to estimate the number of photons due to solar irradiation and compare this to the actual measured photon numbers from the return laser echoes:\\
Solar irradiation at sea level is of order 1 W/m$^2$/nm at $\sim800$ nm but drops by nearly a factor 10x at 1500 nm. This alone is motivation for using longer wavelength sources and thus employing InGaAs SPADs, even if these will typically have  worse detection efficiencies and higher noise levels (dark counts) with respect to silicon based SPADs. Assuming that the detection system is not directly pointed at the sun but rather is collecting light from a screen or wall with e.g. 10\% reflectivity, over an observation radius of 1 cm and through a 1 nm bandwidth filter, we have a reflected power of 0.03 mW at 800 nm (or at 1500 nm with a standard 10 nm bandwidth filter). This power is scattered into the full solid angle and decays as 1/$d^2$. For a detection distance of $d=50$ m and collection through a 1" lens, the actual power on the detector is 2 pW. In order to calculate the number of photons that are actually collected we need to account for the fact that SPAD sensor is not continuously detecting but rather is triggered by the pulsed laser source with a $\sim50$ ns gate time, i.e. for a 1 second acquisition time, triggered by an 80 MHz laser and detection electronics running at 4 MHz (often the actual bottleneck in these applications), the sensor is actually collecting photons for only 0.04 s, corresponding to a collected energy of $4\cdot10^{-13}$ J, i.e. $\sim10^{6}$ photons. With detection efficiencies of the order of 10\%, we can therefore expect $\sim1000$ background photons detected in each time bin (with 1024 time bins in a 50 ns detection window). This number is of the same order of magnitude of the dark count rate for commercial InGaAs SPADs. Experiments show that the number of photons detected from the laser return echo are of order of 1000-10000 photons, distributed in 1-2 time bins: this is therefore sufficient to distinguish the return echo from the background from both the dark counts (thermal noise) and solar irradiation (as long as the detector is not pointed directly at the sun or at a highly reflective/specular surface). These rough estimates confirm that careful choice of laser illumination wavelengths, detector/filter specification and triggering for time-correlated detection, does allow to achieve close to to real-time tracking of moving objects with eye-safe illumination.\\

\subsection{Imaging of light pulse propagation in fibres and slow-light media.}
Light-in-flight photography has been applied to the study of the propagation of laser pulses in dispersive materials. \\
\begin{figure}[t]
\begin{center}
\includegraphics[width=8cm]{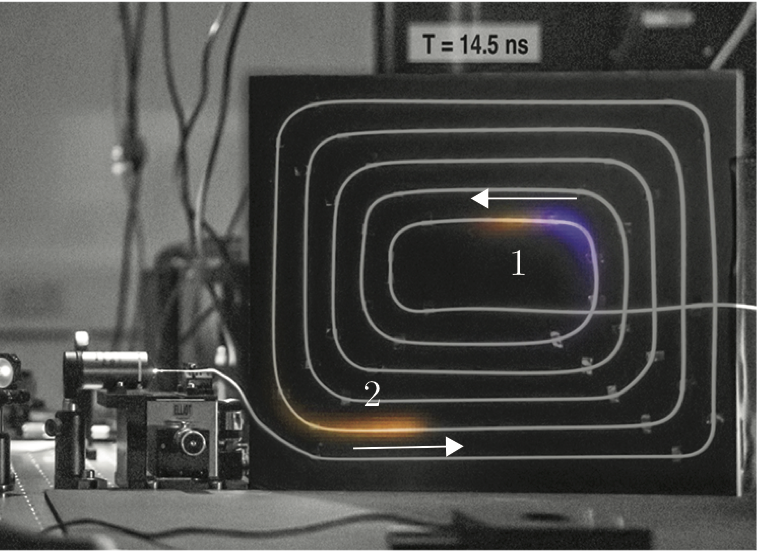}
\caption{Light-in-flight imaging of a femtosecond pulse propagating in a photonic crystal fibre. Two pulses are visible: the first pulse, ``1'', has propagated a further distance and displays a broader spectrum with a strong temporal chirp.}
\label{fiber}
\end{center}
\end{figure}
Femtosecond pulses propagating in photonic crystal fibres undergo a strong transformation as they interact with the glass medium and broaden their spectrum, forming a so-called ``white-light'' pulse \cite{SC}. This process has found many applications and lies at the heart of modern white light sources used in spectroscopy. In Ref.~ \cite{fibre} a few meters of curled photonic crystal fibre were imaged with a SPAD camera. A series of measurements were performed, each time with a different interference filter placed in front of the camera so as to observe the LiF of different spectral components in the pulse. The individual measurements where then combined in a single video after computationally applying a colour corresponding to the filter used. A single frame is shown in Fig.~\ref{fiber} where two pulses from the 80 MHz laser are simultaneously visible: the pulse that has propagated the longest distance in the fibre has also suffered the strongest transformation and exhibits a broader spectrum. The pulse chirp, i.e. the variation of colour along the pulse is also clearly visible. This effect was then used to determine the dispersion characteristics of the fibre.\\
\begin{figure}[t]
\begin{center}
\includegraphics[width=8cm]{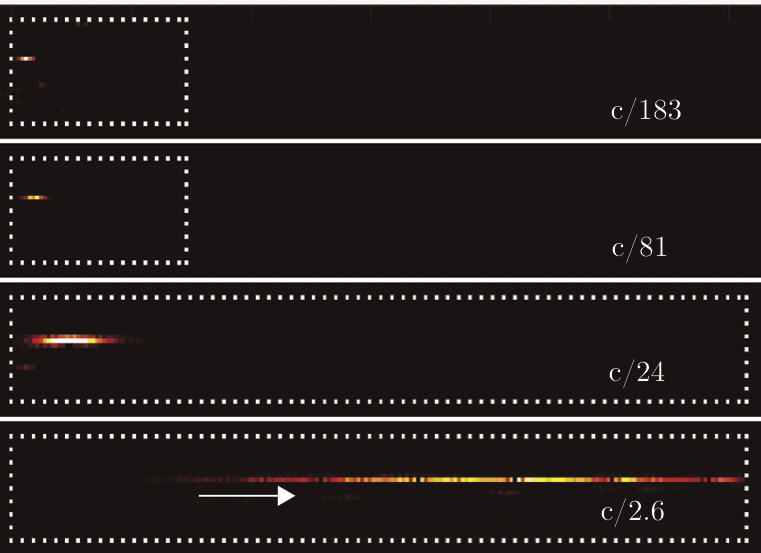}
\caption{Light-in-flight imaging of a nanosecond slow light pulse. Four different speeds are considered, ranging from $c/2.8$ to $c/183$, as indicated in the figure. The increasing pulse spatial shortening with the decreasing group velocity is evident. The dashed boxes indicate the position and shape of the Rb atomic vapour cells: the longer(shorter) is 30(7) cm long. The arrow indicates the laser pulse propagation direction. All measurements are synchronised to the same ``start'' time: the different pulse locations along the propagation direction are a direct manifestation of the slowed propagation. }
\label{slow}
\end{center}
\end{figure}
There are media in which the optical dispersion becomes so large that the group velocity of light is slowed down nearly to a halt \cite{slowlight}. Atomic vapours allow for example to slow light down to less than 100 m/s and even very simple schemes allow to slow light down by factors 100x or more. These systems have been proposed as buffers for light, where information in the form of light pulses can be stored and then released at later times. This becomes also particularly interesting at the single photon level with applications for quantum communication and computation \cite{slow_photon,slow_photon2}. SPAD cameras have been used to directly measure slow light \cite{kali}. The single photon nature of SPAD detection also implies that optical delays of single photons can be characterised. Figure~\ref{slow} shows an example of the same time frame in different settings where light is slowed by different amounts. An interesting observation is the compression of the input pulse from 30 cm (1 nanosecond duration) down to a few mm due to propagation at c/100.
\begin{figure}[t]
\includegraphics[width=8cm]{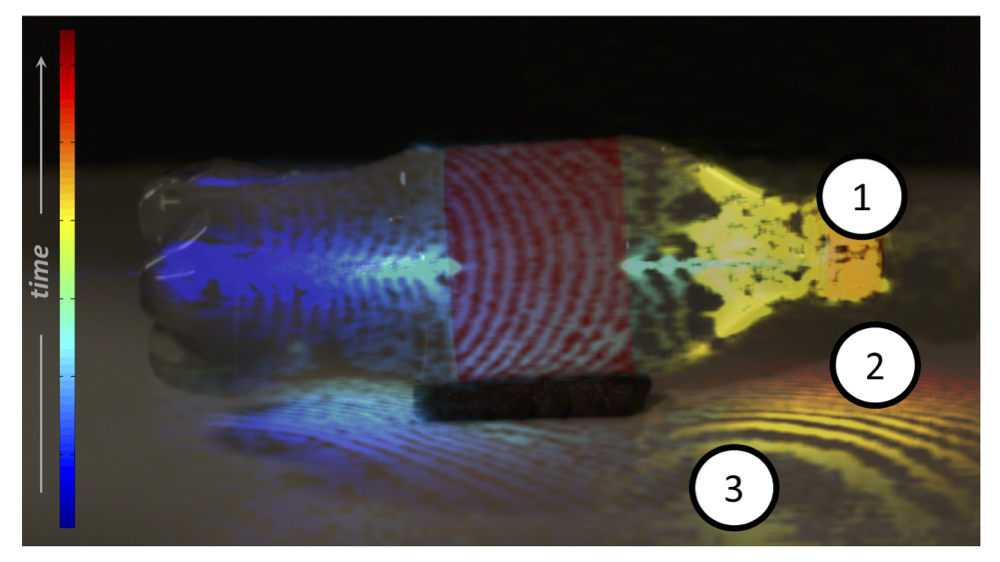}\\
\includegraphics[width=8cm]{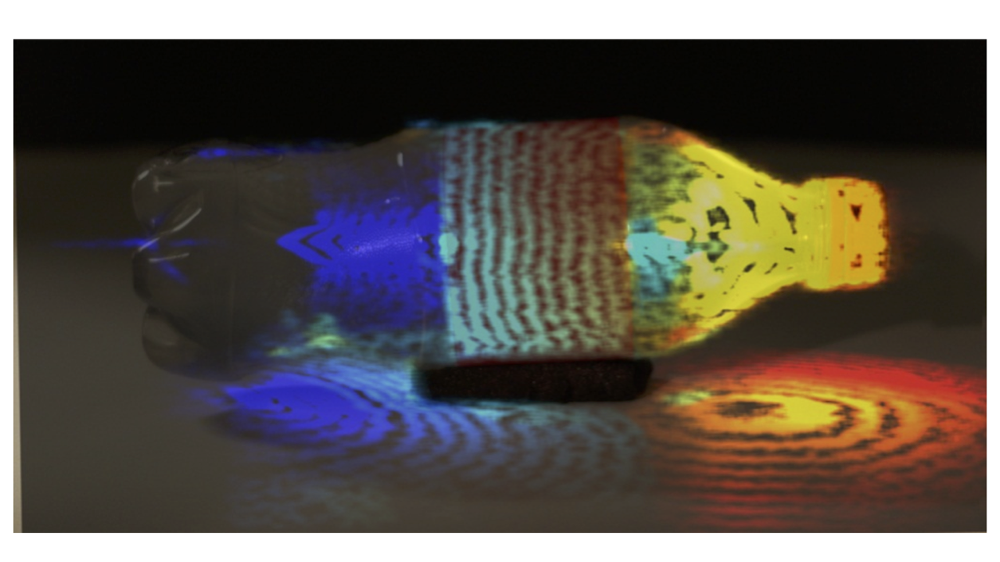}\\
\begin{center}
\caption{Single image visualization of a video of a laser pulse traveling through a soda bottle filled with slightly murky water. Early times are shown in blue, later times in red. The top panel shows the uncorrected video, the bottom panel shows the video after correcting for the travel time from the scene points to the camera. Figure adapted from~\cite{velten_femto-photography:_2013}.}
\label{bottle}
\end{center}
\end{figure}

\subsection{Time of flight distortions and relativistic effects.} 
\begin{figure*}[t]
\begin{center}
\includegraphics[width=16cm]{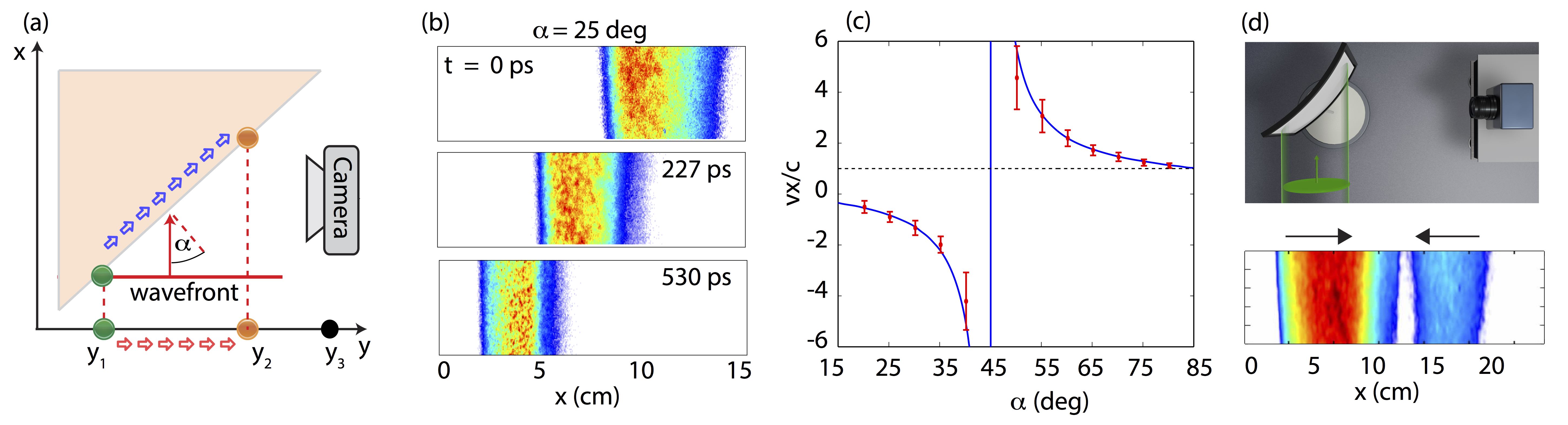}
\caption{(a) interaction geometry showing the angled screen, impinging light wavefront and camera viewpoint. (b) Example of measured wavefront at three different times (as indicated in the figures) and showing the superluminal wavefront scattering point apparently moving from right-to-left when in reality, it is moving in the opposite direction. (c) Full summary of the observed scattering point speeds for varying angles of the observation screen. The solid line shows the theoretical prediction. For angles smaller than 45 deg. the wavefront is  see to propagate backwards. (d) Experimental layout and observed image splitting due to a transition from a sub to a super luminal propagation geometry. Figures are adapted from Ref.~\cite{clerici_super}.}
\label{super}
\end{center}
\end{figure*}
Light-in-flight videos show interesting distortions and artefacts that result from the finite propagation speed of light. The potential to use LiF to record relativistic distortions, in the sense of distortions that are due to the finite speed of light, was first proposed by Duguay et al. \cite{3HSP1971} and then by Abramson \cite{Abramson:83}. Methods have been proposed to account for these effects \cite{Abramson:84v2,Abramson:84,Abramson:85,8HTR1991,velten2013}. LiF photography has more recently also been used to study relativistic effects as opposed to considering these as a distortion that requires correction \cite{9fsPH2007,clerici_super,Laurenzis:16}. \\
 In conventional videos, the speed of light is considered infinite and an event is recorded  in a video at the same time that this occurs in the scene. In light-in-flight videos, the time of flight of the light  from the event to the camera has to be considered. This can lead to strong distortions such as shearing and temporal inversion. Shearing effects were first discussed by Duguay et al. \cite{3HSP1971}: using an optical gating technique with 10 ps resolution \cite{1969gate}, they showed that if two pulses are positioned at different distances from the observer and propagating transversally, the further pulse will appear to be delayed with respect to the closer pulse as a result of the additional time required for light to travel the distance between the two pulses. The authors also point out that a relativistic object with a pattern in its transverse profile would show this pattern when photographed at the speed of light: you would for example be able to see the back of a coin moving transversely at speeds $\sim c$. This effect was indeed verified by Kubota et al. \cite{9fsPH2007} where the Japanese character for ``light'' was impressed on a laser pulse and this was seen to be inverted (as expected when the character is apparently viewed from behind) when imaged with holographic LiF.\\
 Temporal inversion effects have also been observed. These are events in the video that appear to happen at incorrect timings and can also appear in the wrong order. An example from Ref.~\cite{velten2013} is reproduced in Figure~\ref{bottle}. In this video a light pulse hits the cap of a bottle (Event 1). Scattered light travels from there to the floor underneath the cap. The first spot on the floor to be reached by light is the point right underneath the cap (Event 2). Only later, points further away from the cap, but closer to the camera are struck (Event 3). While the events occur in the the order 1 - 2 - 3, the uncorrected video shows them in the order 1 - 3 - 2. This is because Event 3 occurs at a scene that is close to the camera and therefore light from there reaches the camera faster than light from events 1 and 2. This can create effects in the scene that appear to move faster than the speed of light.\\
Later work explicitly addressed the temporal inversion that can occur when imaging an optical wavefront with a technology that can capture this motion \cite{clerici_super}. As with the example described above, the finite propagation speed of light combined with non-trivial observation geometries, can lead to a temporal inversion of events. In particular, a specific effect was originally proposed by Lord Rayleigh: whilst describing the Doppler effect, he noted that if an emitter moves at supersonic speeds,  ``sounds previously excited would be gradually overtaken and heard in reverse of natural order...the observer would hear a musical piece in correct time and tune, but backwards'' \cite{rayleigh}. Due to sound attenuation in air, trying to perform such an experiment and actually hearing a musical piece playing backwards is extremely challenging and, maybe surprisingly, has never been achieved. On the other hand, it is possible to very simply create a scattering geometry for a light wavefront propagating across a wall or screen that provides an effective source of (scattered) light that moves at superluminal speeds. Indeed, as already pointed out by Abramson \cite{abramson}, the intersection point of a plane wave of light intersecting a screen at an angle $\alpha$, moves across the screen at speed $c/\sin\alpha$ [see Fig.~\ref{super}(a)] \cite{Nemiroff3}. Light scattered from this point will therefore give the impression of a moving, superluminal ``source'' of light: by controlling the observation angle of the camera with respect to the screen inclination, it is possible to achieve superluminal propagation along the direction of observation,
\begin{equation}
v_y=\frac{c}{\sin\alpha(\tan\alpha-1)}
\end{equation} 
for $0< \alpha<\pi/4$ thus reproducing the conditions originally laid out by Rayleigh for the case of sound waves and resulting in an apparent negative speed $v_y$. As a result of the temporal inversion of events for superluminal propagation, the wavefront propagating along the inclined screen imaged by a LiF device along the transverse ($x$) plane, will be seen to also propagate with negative speed, i.e. from right to left, even if the wavefront is propagating from left to right. This is shown in Fig.~\ref{super}(b), where for successive propagation times the wavefront, imaged with an iCCD camera, is seen to propagate to the left. Figure~\ref{super}(c) shows the transverse propagation speed $v_x$ as seen by the camera for varying screen angles (red dots), verifying the predicted speed (solid curve):
\begin{equation}
v_x=\frac{c}{1-\cot\alpha}.
\end{equation} 
These results are obtained for constant wavefront speeds (i.e. for a fixed screen angle). It is also possible to curve the screen and study the transition from superluminal to subluminal propagation. The physics of such a transition for a wavefront has been analysed in detail by Nemiroff and Zhong \cite{Nemiroff2,Nemiroff1,Nemiroff4,Zhong:16,Nemiroff5}, with particular emphasis on the implications for astrophysical observations. The prediction is that at the transition, an image pair will be observed: at the transition for super to subluminal, a pair of images is created at the point for which $v_y=c$ whilst image pair annihilation is observed in the opposite case. Figure~\ref{super}(d) shows the experimental layout for observing a sub to superluminal transition and the resulting pair of wavefronts imaged by the camera. The remarkable feature of this measurement is that at all times the wavefront is actually intersecting the screen in one and only point: a pair of wavefronts is observed purely as the result of the finite propagation speed of light and the superluminality of the scattering `source'.\\
While these temporal distortions involve many ideas and concepts that are important to special relativity, the interpretation of a video does not require the transfer between different moving frames of reference and thus is not strictly a relativistic problem. The theory of relativity, however, becomes inescapable when a view with a moving camera has to be rendered. While it is not possible to move the camera at near relativistic speeds, the captured data can be manipulated to show such a scenario using computer graphics techniques. These moving camera renderings require a relativistic rendering model correcting for spatiotemporal distortions and doppler shift~\cite{velten_relativistic_2012}.

\subsubsection{Light-in-flight in astronomy.}
While light-in-flight techniques used on earth typically require fast light sources and detectors, light-in-flight processes can more easily be observed in astronomical phenomena due to the large distances involved. A number of astrophysical sources have been observed in which faster-than-light motion has been inferred from dividing the distance travelled and the arrival times of light features from the object \cite{rees}. One of the main examples are quasars, active galactic nuclei populated by a supermassive black hole. The accretion disk surrounding the black hole  provides matter and energy for the emission of intense jets, containing particles moving at relativistic speeds. 
Quasars are visible only when the jet is directed towards or at a small angle towards the observer: the jets are emitting light at every point along their path, so the emitted light approaches the observer at speeds similar to the jet itself. This causes the light emitted over many years of the jet's travel to actually appear as if  the earliest light emitted and the latest light emitted are separated by a much smaller time period. This is the origin of the illusion of faster-than-light travel, with reports of speeds up to ten times c \cite{10c}. Although there are some controversial examples where the jet angle seems to be too large to explain the measured degree of superluminality, these are probably mainly related to the relative uncertainty in the measurements performed over such large distances on extra-galactic, very high red-shift objects.\\
These superluminal motion measurements were rendered possible due to a new astronomical observation technology, Very Long Baseline Interferometry (VLBI). VLBI allows to perform high precision measurements and, importantly, to track small variations of positions, proper motions, over many years. The astronomical distances involved allow therefore to perform this rather peculiar form of light-in-flight photography by simply extending the measurement baseline of the Earth-based instrument.
\begin{figure*}[t]
\begin{center}
\includegraphics[width=11cm]{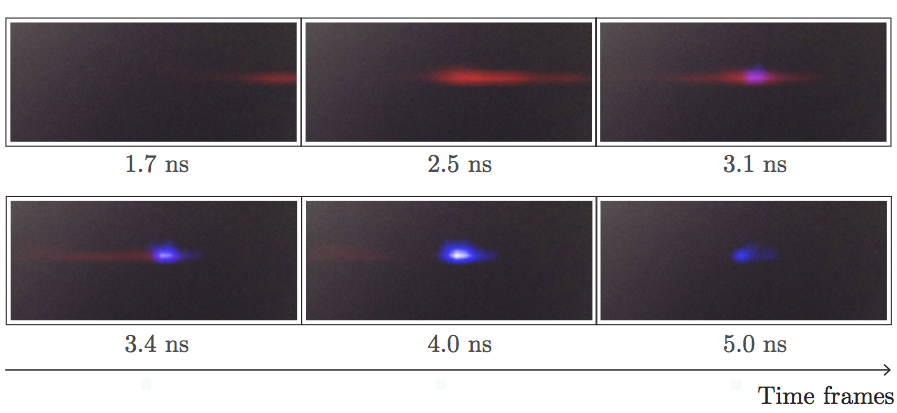}
\caption{Light in Flight photography of an intense femtosecond laser pulse, moving in free space, focusing and creating a plasma through multiphoton ionasation of air molecules. The video is taken with a 32x32 SPAD array. The laser emits pulses at 100 Hz repetition rate, 800 nm wavelength, 90 fs pulse duration and 10 mJ energy/pulse. The images are the superposition of two different measurements, each with a coloured filters to isolate the 800 nm laser light and the $\sim400$ nm, ionised $N_2$ plasma emission. Color is applied to the images in post-processing, corresponding to the two different filters used during the measurements. The absolute time reference provided by the laser pulses allows to precisely temporally align the images. The ionised can be seen to last longer than the excitation light pulse and was measured to have a 600 ps decay time. More details can be found in Ref.~\cite{me1}.}
\label{plasma}
\end{center}
\end{figure*}
\subsection{Imaging of plasma and self focusing.}
Besides imaging the motion of light itself, intensity based light-in-flight cameras can be used to image changes in the scene, as is typically done in conventional videography. Macroscopic objects on human and terrestrial size scales rarely move fast enough to require a camera with light-in-flight resolution. There are however, optical effects in scenes that happen at timescales fast enough to require a light-in-flight camera.
One such example are light filaments created by intense laser pulses self focusing in air. Self focusing is caused by a nonlinear material response present in any medium that causes the index of refraction to vary as a function of the light intensity. A light beam with a gaussian intensity profile can thereby create a varying index profile that acts like a graded index lens and focuses the beam. In air, this self-focusing effect occurs at very high intensities only achievable with ultra-short pulses. A focused pulse continues to collapse until it is intense enough to ionize the air creating a plasma that then defocuses the pulse \cite{Couairon}. It has been suggested, that a filament launched at just the right conditions may propagate for kilometers as a tightly focused beam with interesting implications for remote sensing and energy delivery \cite{kolesik}. Due to the plasma created in the beam path, filaments have been proposed for guiding terahertz \cite{Thz_guide} and microwave radiation \cite{microwave} and can conduct electronic discharges \cite{vidal}, leading to proposals for novel lightning discharge control devices \cite{diels}. Filaments also are sources of terahertz \cite{THz,clerici} and broad band optical radiation \cite{Couairon}, as well as of sound waves \cite{sound}. \\
SPAD arrays have been used to directly image a focusing femtosecond laser pulse and the subsequent evolution of the ionised plasma at the focus. Frames from this video are shown in Fig.~\ref{plasma}, highlighting how SPAD arrays can also be used with low repetition rate lasers (here, just 100 Hz), although this of course comes at the expense of relatively long  exposure times (of order of several minutes) in order to accumulate sufficient photon statistics.\\
A setup similar to the one used in~\cite{velten_femto-photography:_2013} has been used to create movies of light filamentation in air showing light oscillations that may be associated with plasma dynamics~\cite{velten_videos_2015}. In the setup filaments were launched through an aerodynamic window creating a fixed starting point. The light emitted by the filament was imaged by a streak camera. The camera line is scanned across the location of the filament and 200 consecutive lines are arranged to create a video of the filament. 

\section{Future directions and Conclusions.}
Light-in-flight photography was introduced in the late 70's and has gained increased interest in recent years due on the one hand to a series of new optical and digital imaging technologies that have enabled a series of methods for imaging at frame rates of the order of a trillion frames per second. These technologies in turn have enabled a range of applications, many of which still need to reach full maturity and many others are still being uncovered.\\
One example is the use of both spatial and temporal information for improving the ability to image inside diffuse media \cite{allphotons,Boccolini:17}. The final goal will be to provide a new approach to deep imaging inside the human body for medical diagnostics. Significant research is also being invested in the problem of imaging behind occlusions and in general outside the direct line of sight, with large-scale initiatives  funded in the UK and USA. \\
A key aspect to bear in mind is that the renewed interest in LiF and the resulting applications are largely due to innovations in the imaging hardware. It is here therefore that we hope to see more progress in the near future with a faster, more light-efficient detectors being developed. Examples of progress in this sense can be found in the SPAD sensors but research is also continuing in photo-multiplier tubes, with higher temporal resolutions now in the sub-20 ps range \cite{becker} and proposals for new-generation ``streak'' PMTs that could deliver sub-ps timing capability \cite{psPMT}.\\
We conclude with a comment regarding the impact that new technology can have beyond the context in which it was originally conceived. Digressing only slightly, a related example is cinematography and its evolution during the first world war. Before the war, cinema was static and was in essence an extension of a theatrical performance, with a stage and a static viewpoint of the scene. The war however, did not allow any form of staging and introduced the necessity to follow and capture the action as this unfolded. This required the development of new cameras. The unforeseen consequence of this was a radical change in the use of cinematography but also a radically new perception of the world, no longer static but in fast movement and in continuous evolution.  Likewise, the introduction and diffusion of LiF techniques is changing the way in which we visualise our data, perform our measurements and perceive the world. Current research is concentrating on combining these techniques with novel computational approaches that allow to create images in situations that were not previously even considered as a possibility. Full 3D imaging behind a wall has already been demonstrated and work is underway to create cameras that can see through fog, tissue or optical fibres. One wonders, what other exciting applications and developments that we have not even thought of yet might be on the way?\\

\section*{References}

\end{document}